\documentclass[reprint, showpacs, groupedaddress, showkeys, nofootinbib,
 amsmath,amssymb,
 aps,
 natbib,
]{revtex4-1}

\usepackage{graphicx}
\usepackage{dcolumn}
\usepackage{bm}
\usepackage{hyperref}
\usepackage{amssymb}
\usepackage{mathrsfs}
\usepackage{amsmath,fleqn}

\def\JPhysA{J. Phys. A: Math. Gen.}
\def\EPJST{Eur. Phys. J. Spec. Top.}
\def\EPL{Europhys. Lett.}

\def\PRA{Phys. Rev. A}
\def\PRE{Phys. Rev. E}
\def\PRL{Phys. Rev. Lett.}
\def\PLA{Phys. Lett. A}

\def\PTP{Prog. Theor. Phys.}
\def\LNP{Lect. Notes Phys.}

\newcommand{\bea}{\begin{eqnarray}}
\newcommand{\eea}{\end{eqnarray}}
\newcommand{\ba}{\begin{array}}
\newcommand{\ea}{\end{array}}

\newcommand{\imag}{{\rm i}}
\newcommand{\be}{\begin{equation}}
\newcommand{\ee}{\end{equation}}
\newcommand{\bt}{\begin{teo}}
\newcommand{\et}{\end{teo}}

\begin{document}

\preprint{APS/123-QED}

\title{Survey on the role of accelerator modes for anomalous diffusion:\\ The case of the standard map}

\author{Thanos Manos}
\email{thanos.manos@uni-mb.si}
\affiliation{CAMTP - Center for Applied Mathematics and Theoretical Physics, University of Maribor, Krekova 2, SI-2000 Maribor, Slovenia}
\affiliation{School of Applied Sciences, University of Nova Gorica, Vipavska 11c, SI-5270 Ajdov\v s\v cina, Slovenia.}

\author{Marko Robnik}
\email{Robnik@uni-mb.si}
\affiliation{%
 CAMTP - Center for Applied Mathematics and Theoretical Physics, University of Maribor, Krekova 2, SI-2000 Maribor, Slovenia.
}

\date{\today}

\begin{abstract}

We perform an extensive and detailed analysis of the generalized diffusion processes in deterministic area preserving maps with noncompact phase space, exemplified by the standard map, with the special emphasis on understanding the anomalous diffusion arising due to the accelerator modes. The accelerator modes and their immediate neighborhood undergo ballistic transport in phase space, and also the greater vicinity of them is still much affected (``dragged'') by them, giving rise to the non-Gaussian (accelerated) diffusion. The systematic approach rests upon the following applications: the GALI method to detect the regular and chaotic regions and thus to describe in detail the structure of the phase space, the description of the momentum distribution in terms of the L\'evy stable distributions, the numerical calculation of the diffusion exponent and of the corresponding diffusion constant. We use this approach to analyze in detail and systematically the standard map at all values of the kick parameter $K$, up to $K=70$. All complex features of the anomalous diffusion are well understood in terms of the role of the accelerator modes, mainly of period 1 at large $K\ge 2\pi$, but also of higher periods (2,3,4,...) at smaller values of $K\le 2\pi$.
\end{abstract}

\pacs{05.45.Pq, 05.45.Ac, 05.60.Cd}
\keywords{Standard map, anomalous diffusion, accelerator modes, L\'evy stable distribution, regular and chaotic motion}
\maketitle


\section{Introduction \label{intro}}
The question of transport in Hamiltonian systems goes back to the early works by \citet{C79}, \citet{RechWhi1980PRL}, \citet{RechRosWhi1981PRA}, \citet{CarMeiBha1981PRA}, \citet{MCGCKA1983PhyD}, \citet{Kar1983PhyD}, \citet{HorHatIshMor1990PTP,IshHorKobMor1991PTP,OucMorHorMor1991PTP,MorOkaTom1991PTP}, \citet{MacMeiPer1984aPRL,MacMeiPer1984bPhyD}, \citet{Zas2007} (and the references therein), \citet{Stef_etal1998PRE} and many others. In particular, it has been shown that in cases of sufficiently strong chaoticity the transport can be diffusive, in the sense of exhibiting normal diffusion, where the distribution of the relevant quantity, e.g. angular momentum, in the phase space is Gaussian with the variance growing linearly with time, characteristic of stochastic diffusion processes and random walks. In the Chirikov map \cite{C79}, which is describing the classical kicked rotator, called also standard map [see the Eq.~(\ref{SM2}) below], such normal diffusion has indeed been observed in the very early days.  This finding can be immediately understood by realizing that once the jumps in the value of (rotation) angle $\theta$ are big enough, the increments of (angular) momentum $P$ become essentially uncorrelated and thus random, resulting in a Gaussian or Brownian random walk. This is quite easy to understand qualitatively and also to calculate the approximate diffusion constant as a function of the kick parameter $K$, which turns out to be approximately $D_1=K^2/2$, with the definition in Eq.~(\ref{varp}) below, and this estimate is valid for sufficiently large $K$, but not too large. Namely, as found by \citet{RechWhi1980PRL,RechRosWhi1981PRA}, and using the Fourier transform technique introduced by \citet{Aba1981PhyD,AbaCraw1981PLA} in a refined theory, there are substantial corrections to this simple estimate and the diffusion constant as a function of $K$ displays well known and well understood oscillations, approximately described in Eq.~(\ref{Dcl}) below. For a review see \cite{LL1992,Rei2004,ZazEdeNiy1997Chaos,Zas2007}.

However, even this picture, not easy to derive theoretically, is too simple, as there are intervals of $K$ in which the diffusion is not normal, but is instead anomalous, mainly superdiffusion, such as described in Eq.~(\ref{varp}) below with $\mu > 1$. The reason lies in the phenomenon observed and correctly interpreted first by \citet{C79}, called accelerator modes: within some intervals of $K$  at $K\ge 2\pi$ there exist stable (regular) regions (islands) in the phase space surrounding the periodic orbits of period 1 in the compact phase space, corresponding to the jumps in $P$ equal to $2\pi$ or integer multiples of $2\pi$. If we de-compactify the phase space, making it an infinite cylinder, we observe infinite transport along the cylinder. The orbits inside such accelerator regions are thus simply ballistically (linearly in time) transported along the cylinder, in both directions $P\rightarrow\pm \infty$. Thus, all orbits trapped inside the accelerator modes display ballistic diffusion with the diffusion exponent $\mu=2$. Moreover, those orbits that either originate from the neighborhood of the accelerator modes, or come close to them in the course of time, becoming trapped there for a while
due to the stickiness of the neighborhood (containing cantori), get ``dragged'', or accelerated by them, and therefore display anomalous diffusion with $\mu >1$ (superdiffusion). Furthermore, apart from the accelerator modes of period 1, there exist also accelerator modes of higher periods, 2,3,4..., which we also observe in this work, but their role becomes less important with increasing $K$ much faster than for period 1.

There have been many attempts to account for the anomalous diffusion in the area preserving maps, most notably by \citet{Veneg2007PRL, Veneg2008PRE,Veneg2008PRL}, but it seems still largely impossible to predict the diffusion exponent $\mu$ (see also \cite{ZasEde2000Chaos}) for a set of initial conditions at a given $K$.

In this paper we study the diffusion properties in area preserving maps, exemplified by the standard map of Chirikov. This work is actually motivated by our extensive study \cite{ManRob2013PRE,BatManRob2013} of the quantum kicked rotator introduced by \citet{CCFI79}, in which - at the semiclassical level - it is necessary to understand in detail the classical diffusion, in order to set up a theory of (exponential) quantum (or dynamical) localization. Our previous work was stimulated by the series of pioneering and classic papers by \citet{Izr1988,Izr1989,Izr1990}. Thus, we set out to understand in detail the diffusion process for all $K$ in the interval $0\le K \le 70$ which encompasses  a sufficiently large interval, where rich dynamics with islands of stability, weak and/or strong chaotic seas and effects due to accelerator modes still play a significant role. In this sense, we calculate at each $K$ the \textit{diffusion exponent} $\mu$, the corresponding \textit{diffusion constant} $D_{\mu}$ and the parameter $\alpha$ of the relevant \textit{L\'evy stable distribution} in the case of non-Gaussian (anomalous, super-) diffusion.

The role of accelerator modes in anomalous diffusion has been already discussed broadly in all the articles mentioned before. However, in almost all cases, the authors study these phenomena in a rather local way. In more detail, they focus either on the effect of the isolated accelerator modes, by studying the diffusion properties of ensembles evolved under their effect, or sometimes together with a sample of nearby chaotic initial conditions. Here, we attempt a systematic approach by studying the detailed diffusion properties for a great variety of ensembles of initial conditions.

Being motivated originally by the quantized standard map, and, in order to associate the above classical transport properties with the quantum characteristic time scales (e.g. Heisenberg and localization time), we restrict the upper limit of the final number of iteration to the order of few thousands (in most of the cases up to $n=5000$). Moreover, we do this not only for an ensemble of initial conditions covering the entire phase space, but also locally, for small cells on a fine grid. By doing this, we reveal interesting structures in the phase space, all of them directly correlated with the degree of chaos as detected and measured by the Generalized ALignment Index (GALI) method, and we claim to understand them in detail in terms of the accelerator modes of period 1, and also of periods 2,3,4,... Hence, we manage to resolve anomalous diffusion even in tiny regions of the phase space by quantifying the different degree of the local and global diffusion when the kick parameter $K$ varies. In this way we classify the different kinds of stable regions according to the different transport processes which are associated to islands of stability or/and accelerator modes.

The paper is structured as follows: In Sec.~\ref{sec:model} we define and describe briefly the model (standard map), in Sec.~\ref{sec:methods} we present the methods of analysis, namely (i) the GALI method for the accurate and detailed distinction between chaotic and regular motion in the model's phase space and (ii) the L\'evy stable distributions for the study of the diffusive variable momentum. In Sec.~\ref{sec:res} we present our main results on the dynamical effect of accelerator modes on the diffusion exponent and the $\alpha$-L\'evy parameter. Finally, in Sec.~\ref{sec:conc} we summarize and conclude the main findings of this work.

\section{The model \label{sec:model}}
One of the main models of time-dependent systems is the kicked rotator introduced by \citet{CCFI79}. We introduce it here in detail for the purpose of defining and fixing the variables and the notation. The Hamiltonian function is
\be  \label{KR}
H= \frac{p^2}{2I} + V_0 \,\delta_T(t)\,\cos \theta.
\ee
It is one of the most important paradigms of classical conservative (Hamiltonian) systems in nonlinear dynamics. Here $p$ is the (angular) momentum, $I$ the moment of inertia, $V_0$ is the strength of the periodic kicking, $\theta$ is the (canonically conjugate, rotation) angle, and $\delta_T(t)$ is the periodic
Dirac delta function with period $T$. Since between the kicks the rotation
is free, the Hamilton equations of motion can be immediately integrated,  and thus the dynamics can be reduced to the standard mapping, or so-called Chirikov-Taylor mapping, given by
\be \label{SM1}
 \left\{
 \begin{aligned} \label{SM1}
p_{n+1} &= p_n + V_0 \sin \theta_{n+1},\\
\theta_{n+1} &= \theta_n + \frac{T}{I} p_n,
 \end{aligned}
 \right.
\ee
and introduced in \cite{T69,F72,C79}. Here the quantities $(\theta_n, p_n)$ refer to their values just immediately after the $n$-th kick. Then,
by introducing new dimensionless momentum $P_n = p_nT/I$, we get
\be
 \left\{
 \begin{aligned} \label{SM2}
P_{n+1} &= P_n + K \sin \theta_{n+1},\\
\theta_{n+1} &= \theta_n  + P_n,
 \end{aligned}
 \right.
\ee
where the system is now governed by a single classical {\em dimensionless}
kick parameter $K=V_0 T/I$, and the mapping is area preserving.

The generalized diffusion process of the standard map [Eq.~(\ref{SM2})] is defined by
\be \label{varp}
\langle(\Delta P)^2\rangle = D_{\mu}(K) n^{\mu},
\ee
where $n$ is the number of iterations (kicks), and the exponent $\mu$ is in the interval $[0,2)$, and all variables $P$, $\theta$ and $K$ are dimensionless. Here $D_{\mu}(K)$ is the generalized classical diffusion constant. In the case $\mu=1$ we have the \textit{normal diffusion}, and $D_1(K)$ is then the normal diffusion constant, whilst in the case of anomalous diffusion we observe \textit{subdiffusion} when $0 < \mu < 1$  or \textit{superdiffusion} if $1 <\mu \le2$. In the case $\mu=2$ we have the \textit{ballistic transport} which is associated strictly with the presence of accelerator modes.

In the case of the normal diffusion $\mu=1$ the theoretical value of $D_1(K)$ is given in the literature, e.g. in \cite{Izr1990} or \cite{LL1992},
\begin{flalign} \label{Dcl}
 D_{1}(K)=
\begin{cases}
 \frac{1}{2} K^2\left [1- 2J_2(K) \left (1-J_2(K) \right ) \right ], \text{if} \ K \ge 4.5 \\
 0.15(K-K_{cr})^3, \text{if} \ K_{cr} < K \le 4.5
\end{cases},
\end{flalign}
where $K_{crit}\simeq 0.9716$ and $J_2(K)$ is the Bessel function. Here we neglect higher terms of order $K^{-2}$. However, there are many important subtle details in the classical diffusion further discussed below.

\begin{figure}\centering
\includegraphics[width=\columnwidth]{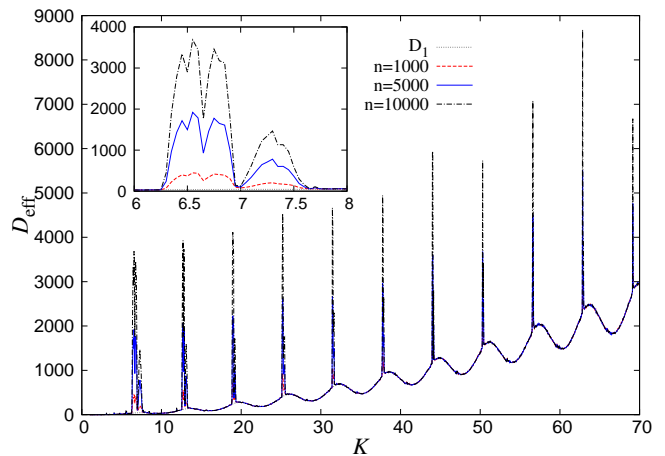}
\caption{(Color online) The classical diffusion constant  $D_{\rm eff}=\langle(\Delta P)^2\rangle/n$ for the standard map as a function of $K$ ($\delta K=0.05$) for three discrete times $n$, i.e., the number of the iterations of the standard map, $n=1000$ (lower red dashed line), $n=5000$ (intermediate blue solid) and $n=10000$ (upper black dot-dashed). In the background we have plotted the classical diffusion constant $D_1$ (gray dotted line) [Eq.~(\ref{Dcl})]. The presence of  accelerator modes at certain intervals of $K$ (and the sticky objects around) generate anomalous diffusion which is rendered by peaks. Here we used $\approx$100000 $(314 \times 314$) initial conditions uniformly distributed in a grid on the entire phase space $[0,2\pi]\times[0,2\pi]$.}
\label{figDcl}
\end{figure}

The dependence of the diffusion constant for the growth of the variance of the momentum on $K$ is very sensitive, and described in the theoretical result [Eq.~(\ref{Dcl})], and fails around the period 1 accelerator mode intervals
\be \label{acmdint}
(2\pi n) \le K \le \sqrt{(2\pi n)^2 +16 },
\ee
$n$ any positive integer. In these intervals for the accelerator modes $n=1$ we have two \textit{stable fixed points} located at $p=0,\; \theta = \pi -\theta_0$ and $p=0,\; \theta = \pi +\theta_0$, where  $\theta_0 = \arcsin (2\pi/K)$. There are two \textit{unstable fixed points} at $p=0,\; \theta = \theta_0$ and $p=0,\; \theta = 2\pi - \theta_0$. For example, in the case $K=6.5$ we have $\theta_0 \approx 1.31179$. Moreover, as the diffusion might even be anomalous, we have recalculated the \textit{effective} diffusion constant $D_{\rm eff}=\langle(\Delta P)^2\rangle/n$ numerically, which in general is not equal to the $D_{\mu}$ defined in Eq.~(\ref{varp}). In Fig.~\ref{figDcl} we show the $D_{\rm eff}$ for the standard map as a function of $K$ for three discrete times $n$, i.e., the number of the iterations of the standard map, $n=1000$ (lower red dashed line), $n=5000$ (intermediate blue solid) and $n=10000$ (upper black dot-dashed). In the background we have plotted the theoretical diffusion constant $D_1$ taking into account only the normal diffusion (gray dotted line) [Eq.~(\ref{Dcl})]. The presence of  accelerator modes at certain intervals of $K$ (and the sticky objects around) generates anomalous diffusion which is rendered by peaks. Here we used $\approx$ 100000 ($314 \times 314$) initial conditions uniformly distributed in a grid on the entire phase space $[0,2\pi]\times[0,2\pi]$. We see that the dotted theoretical curve stemming from Eq.~(\ref{Dcl}) describes the diffusion constant well outside the accelerator mode intervals. In general, however, the diffusion might be non-normal, described in Eq.~(\ref{varp}). There are also accelerator modes of higher period (2,3,4...) observed and examined in this paper.

\begin{figure}\centering
\includegraphics[width=\columnwidth]{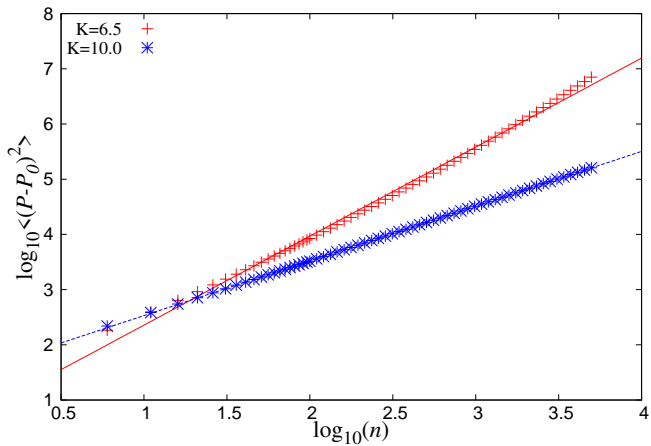}
\caption{(Color online) The variance of the momentum $P$ in the standard map [Eq.~(\ref{SM2})] with $K=6.5$ (red crosses) where small islands and accelerator mode of period 1 are present and $K=10.0$ (blue stars) where the phase space is fully chaotic for the same initial conditions as in Fig.~\ref{figDcl} as a function of the discrete time $n$ (number of iterations), in log-log representation. The two slopes associated with different types of diffusion are $\mu(K=6.5)=1.61252$ (dotted), $\mu(K=10.0)=0.991334$ (solid) with standard deviation errors $\pm$0.01271 (0.7881$\%$) and $\pm$0.0009537 (0.0962$\%$) respectively.}
\label{figDcln}
\end{figure}

In Fig.~\ref{figDcln} we show the variance of the momentum $P$ in the standard map [Eq.~(\ref{SM2})] with $K=6.5$ (red crosses) where small islands and accelerator mode of period 1 are present and $K=10.0$ (blue stars) where the phase space is fully chaotic for the same initial conditions as in Fig.~\ref{figDcl} as a function of the discrete time $n$ (number of iterations), in log-log representation. The two slopes associated with different types of diffusion are $\mu(K=6.5)=1.61252$ (dotted), $\mu(K=10.0)=0.991334$ (solid) with standard deviation errors $\pm$0.01271 (0.7881$\%$) and $\pm$0.0009537 (0.0962$\%$) respectively.

\section{Methods of analysis}\label{sec:methods}

\subsection{The GALI method \label{sec:GALIsection}}
Let us consider a discrete time $t=n \in \mathbb{N}$ conservative dynamical system defined by a $2N$-dimensional ($2N$D) symplectic map $F$. The evolution of an orbit in the $2N$D space $\mathcal{S}$ of the map is governed by the difference equation
\be
\mathbf{x}(n+1)\equiv\mathbf{x}_{n+1}=F(\mathbf{x}_n).
\label{eq:map_gen}
\ee
In this case, the evolution of a deviation vector $\mathbf{w}(n)\equiv \mathbf{w}_n$, with respect to a reference orbit $\mathbf{x}_n$, is given by
the corresponding tangent map
\be
  \mathbf{w}(n+1)\equiv \mathbf{w}_{n+1}=
  \frac{\partial F}{\partial\mathbf{x}} (\mathbf{x}_n)\cdot\mathbf{w}_n.
\label{eq:w_map}
\ee
For $2N$D maps (and $N$ degrees of freedom flows) the Generalized Alignment Index of order $k$ (GALI$_k$), $2 \leq k \leq 2N$, is determined through the evolution of $k$ initially linearly independent deviation vectors $\mathbf{w}_k(0)$. To avoid overflow problems, the resulting deviation vectors $\mathbf{w}_k(t)$ are continually normalized, but their directions are kept intact. Then, according to \citet{SBA:2007} GALI$_k$ is defined as the volume of the $k$-parallelogram having as edges the $k$ unit deviation vectors $\hat{\mathbf{w}}_i(t)=\mathbf{w}_i(t)/ \|\mathbf{w}_i(t) \|$, $i=1,2,\ldots,k$, determined through the wedge product of these vectors as
\be
{\rm GALI}_k(t)=\| \hat{\mathbf{w}}_1(t)\wedge \hat{\mathbf{w}}_2(t)\wedge \cdots
\wedge\hat{\mathbf{w}}_k(t) \|,
\label{eq:GALI}
\ee
with $\| \cdot \|$ denoting the usual norm. From this definition it is evident that if at least two of the deviation vectors become linearly dependent, the wedge product in Eq.~(\ref{eq:GALI}) becomes zero and the GALI$_k$ vanishes. The GALI method is a generalization of the Smaller ALignment Index (SALI) introduced in \cite{SkoJPhA2001} while practically the GALI$_2$ is equivalent to the SALI which also requires two deviation vectors for its calculation \cite{SBA:2007}.

The behavior of GALI$_k$ for regular and chaotic orbits was theoretically studied in \cite{SBA:2007,SkoBouAntEPJST2008}, where it was shown that all GALI$_k(t)$ tend exponentially to zero for chaotic orbits, with exponents that depend on the first $k$ Lyapunov exponents of the orbit \cite{Ben1980Mecc,SkoLNP2010}. In the case of regular orbits, GALI$_k$ remains practically constant and positive if $k$ is smaller or equal to the dimensionality of the torus on which the motion occurs, otherwise, it decreases to zero following a power law decay. In the particular case of 2D maps the GALI$_2$ tends to zero both for regular and for chaotic orbits, following however completely different time rates (exponential \textit{vs}. power law), which again allows us to distinguish between the two cases \cite{SkoJPhA2001}, as explained below.

Before studying the global dynamics of the map (\ref{SM2}) let us examine in more detail the behavior of GALI$_2$ for regular and chaotic orbits of a 2D map. In the case of a chaotic orbit, any two deviation vectors will be aligned to the direction defined by the largest Lyapunov exponent $\lambda_1$, and consequently GALI$_2$ will tend to zero following an exponential decay of the form SALI/GALI$_2 \propto e^{-2\lambda_1 n}$, with $n$ being the number of iterations \cite{SkoAntBouVra2004JPhA}. In the case of regular orbits any two deviation vectors tend to fall on the tangent space of the torus on which the motion lies. For a 2D map this torus is an 1D invariant curve, whose tangent space is also 1D and consequently any two deviation vectors will become aligned. Thus, even in the case of regular orbits in 2D maps the GALI$_2$ tends to zero. This decay follows a power law \cite{SkoJPhA2001} having the form GALI$_2 \propto 1/n^2$ \cite{SBA:2007}.

\begin{figure}\centering
\includegraphics[width=\columnwidth]{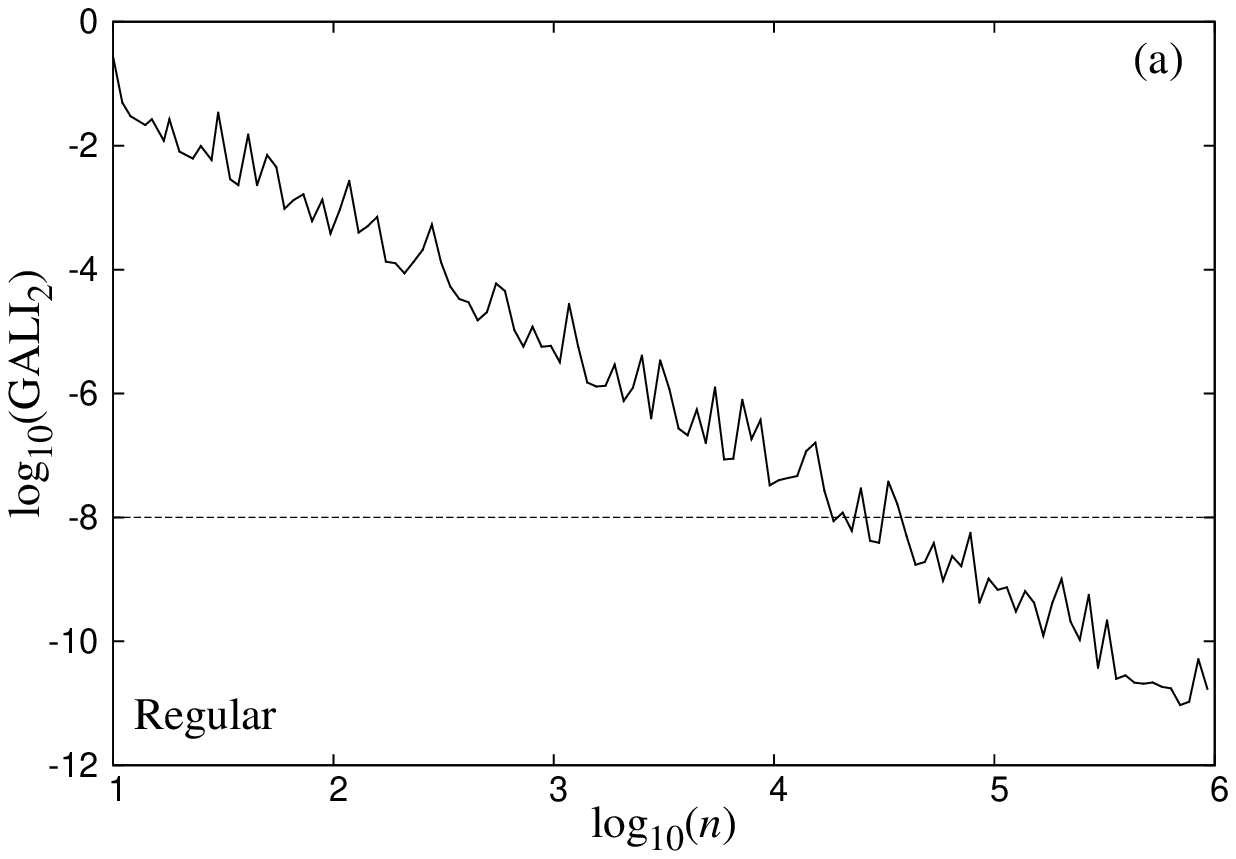}
\includegraphics[width=\columnwidth]{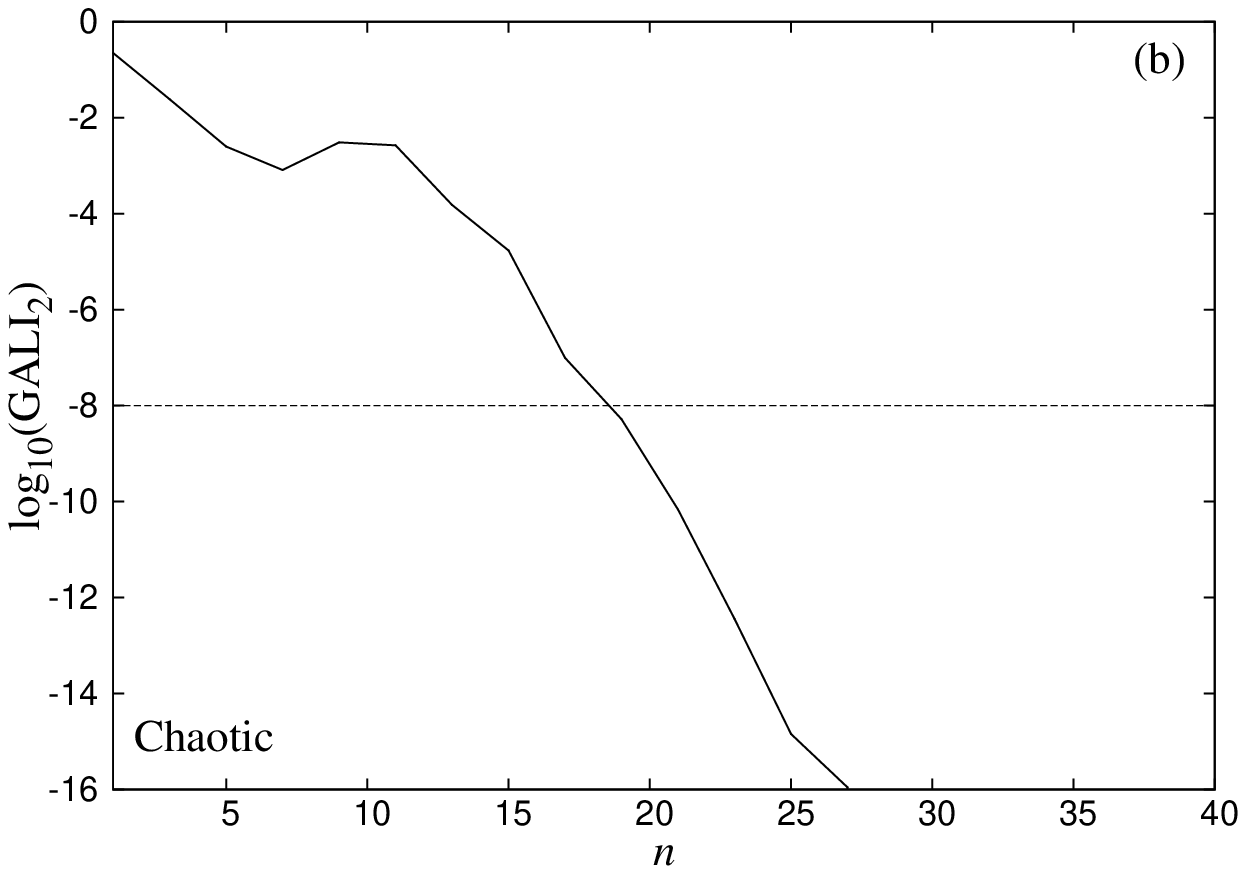}
\caption{The evolution of GALI$_2$ for (a) the regular orbit with initial condition $(\theta,P)=(3.5,0.0)$ and (b) the chaotic orbit with initial condition $(\theta,P)=(1.0,0.0)$ of the standard map [Eq.~(\ref{SM2})] for $K=3.1$, with respect to the number of iterations $n$.}
\label{figGALIorb}
\end{figure}

It is exactly this different behavior of the index that allows us to use GALI$_2$ for a fast and clear distinction between regions of chaos and order in the 2D phase space of the standard map. From the results of Fig.~\ref{figGALIorb}(a), we see that only after $n >10000$ iterations the value of GALI$_2$ of a regular orbit becomes of the order of $10^{-8}$, while for a chaotic orbit [Fig.~\ref{figGALIorb}(b)] the GALI$_2$ has already reached extremely small values (only after $\approx 20$ iterations). Thus, the percentage of chaotic orbits for a given value of $K$ can be computed as follows: We follow the evolution of orbits whose initial conditions lie on a 2D grid of $500\times 500$ equally spaced points on the 2D phase space of the map [dividing in this way the $(\theta,P)$ plane in squares] and register for each orbit the value of GALI$_2$ after $n=50$ iterations. All orbits having values of GALI$_2$ significantly smaller than $10^{-8}$ at $n=50$ are characterized as chaotic while all others are considered as non--chaotic.

The aforementioned threshold $10^{-8}$ has been broadly used and explained in previous works (see for example (see e.g. \cite{MSAB2008,MSB2008,MSB2009}) as a rather efficient threshold. Let us point out that the $10^{-8}$ is in practice the ``zero'' for single precision while the  $10^{-16}$ for double. The results regarding the chaotic or regular nature of a trajectory under study are not so sensitive to the chosen threshold and they still stand if it is varied. It has been tested thoroughly that, in this system, the $n=50$ iterations is sufficiently long for the distinction between the exponential (chaotic motion) or power law (regular motion) decay.

As for the classical system [Eq.~(\ref{SM2})], we mention that the fraction of the regular part of the classical phase space has been systematically explored using the SALI/GALI method for the distinction between chaotic and regular classical motion and its quantification for simple (and even for coupled) standard map(s) (see \cite{MSAB2008,MSB2008,MSB2009} and references therein), showing that this fraction decreases with $K$ following the power laws found by \citet{DFFC2003,CHDF2005IJBC}. However, there are important subtleties about the classical diffusion process and $D_{\mu} (K)$ which we now discuss.

\subsection{L\'evy stable distribution}\label{sec:Levy}
The physical origin and relevance of the L\'evy stable distribution to this kind of problems, like the standard map, is well summarized in e.g. \citet{Zas2007}, \citet{ZazEdeNiy1997Chaos}, \citet{KlaZum1994PRE}, \citet{GeiNieZac1985PRL} and \citet{ZasEde2000Chaos}. In probability theory, an $\alpha$-L\'evy skew stable distribution is a four parameter family of continuous probability distributions, characterizing the location, scale, skewness and kurtosis. Following \citet{nolan:2013}, for a random variable $X$ with distribution function $F(x)$, the characteristic function is defined by $\phi(u)=E \exp(\imag u X)=\int_{-\infty}^{\infty} \exp(\imag u x) dF(x)$. Then, a random variable $X$ is \textit{stable} if and only if $X\overset{\delta}{=}aZ+b$, with $a>0$, $b \in \mathbb{R}$ and $Z$ is a random variable with characteristic function
\begin{flalign} \label{Zcf}
 E \exp(\imag u Z) =
  \begin{cases}
  e^{-|u|^{\alpha}[1-\imag \beta \tan \frac{\pi \alpha}{2}({\rm sign}u)]},  \ & \alpha \neq 1 \\
  e^{-|u|[1+\imag \beta \tan \frac{2}{\pi}({\rm sign}u)] \log|u|}, \ & \alpha = 1 \end{cases},
\end{flalign}
where $0 < \alpha \le 2$ and $-1 \le \beta \le 1$ (the symbol $\overset{\delta}{=}$ indicates that both expressions have the same probability law). We then adopt the parametrization\footnote{There are several parameterizations notated by different $k$ for stable laws \cite{nolan:2013} which originate from the study of different problems at different historical periods. The one with $k=0$ is considered to be the best choice for numerical computations having the simplest form for the characteristic function and being continuous in all four parameters.} $k=0$, $S(\alpha,\beta,\gamma,\delta;0)$ for which the random variable $X$ given by
\begin{flalign}
  X \overset{\delta}{=}
 \begin{cases}
  \gamma(Z-\beta \tan(\frac{\pi \alpha}{2}))+\delta,  \ & \alpha \neq 1 \\
  \gamma Z + \delta, \ & \alpha = 1
\end{cases},
\end{flalign}
has characteristic function
\begin{flalign}
 & S(\alpha,\beta,\gamma,\delta;0) \equiv E \exp(\imag u X) = \\ \nonumber
& \begin{cases}
  e^{\imag u \delta - \gamma^{\alpha}|u|^{\alpha}(1+\imag \beta(-1+|u \gamma|^{1-\alpha}){\rm sign}(u)  \tan(\frac{\pi \alpha}{2}))},  \ & \alpha \neq 1 \\
  e^{\imag u \delta - \frac{\gamma |u|(\pi+2\imag \beta \log(|u \gamma|)) {\rm sign}(u)}{\pi}}, \ & \alpha = 1
\end{cases},
\end{flalign}
where $Z=Z(\alpha,\beta)$ is defined as described in Eq.~(\ref{Zcf}), $\alpha \in (0,2]$ is the index of stability or characteristic exponent, $\beta \in [-1,1]$ the skewness parameter, $\gamma >0$ the scale parameter and $\delta \in \mathbb{R}$ location parameter. For the fits with data we used the \textit{Stable Distribution} package of \textit{Mathematica} \cite{RimNol2005MathJ}. Two important special cases are the Gaussian distribution with $\alpha =2$ and the Cauchy-Lorentz with $\alpha =1$ which are the only ones with an explicit closed formula.

\section{Results: the dynamical effect of accelerator modes on the diffusion exponent and the $\alpha$-L\'evy parameter.} \label{sec:res}

In this section we discuss the role of the accelerator modes in the local and global dynamics of the phase space of the standard map [Eq.~(\ref{SM2})]. In more detail, we draw our attention on the way they affect the diffusion process and also how they are reflected in the momenta probability distribution.

\begin{figure*}\centering
\includegraphics[width=15cm,height=8cm]{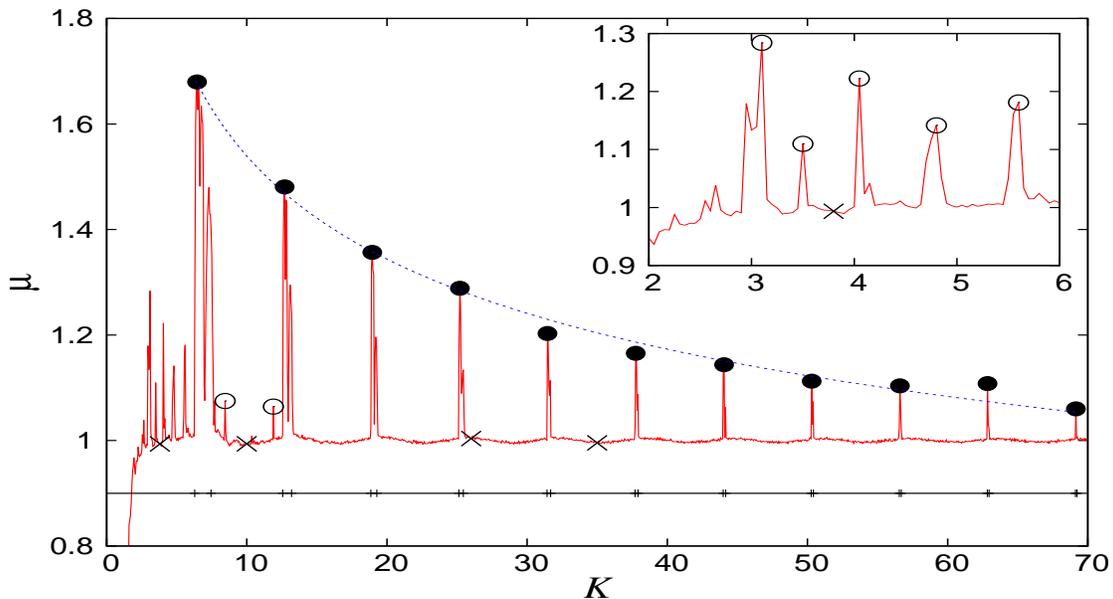}
\caption{(Color online) The diffusion exponent $\mu$ as a function of $K$  after $n=5000$ iterations and for $\approx$100000 $(314\times 314)$ initial conditions on the plane $(\theta,P)=(0,2\pi)$. The intervals on the black horizontal line $\mu=0.9$ indicate the intervals of stable accelerator modes of period 1 [Eq.~(\ref{acmdint})]. All intervals of $K$ with exponent $\mu \approx 1$ are associated with normal diffusion processes. The large peaks (appearing mainly for $K > 2\pi$ marked with full black circles) reflect the anomalous diffusion accelerator modes (mainly of period 1). The smaller peaks for $K < 2\pi$ (more clearly presented in the inset panel) originate by accelerator modes of higher period together with those for $2\pi < K < 4\pi$ marked with empty circle and a few typical examples close to those peaks [marked with the symbol ($\times$)], for which the diffusion is normal, are studied thoroughly later on. The blue dotted line corresponds to the power law which describes the decay of the exponent $\mu$ of the main peaks' amplitude due to accelerator modes of period 1 (see text for more details).}
\label{figKvsmu}
\end{figure*}

In Fig.~\ref{figKvsmu} we show the diffusion exponent $\mu$ as a function of $K$ after $n=5000$ iterations, using a fine grid of $314\times 314~(\approx 100000)$ initial conditions on the plane $(\theta,P)=(0,2\pi)$. The $\mu$ is calculated by the slopes, of the lines of the variance of the momentum $P$ as a function of iterations, as it is described in Sec.~\ref{figDcln} [Eq.~(\ref{varp})] and for a grid of cells on the entire phase space. The intervals on the black horizontal line $\mu=0.9$ indicate the intervals of stable accelerator modes of period 1 [Eq.~(\ref{acmdint})]. All intervals of $K$ with exponent $\mu \approx 1$ are associated with normal diffusion processes. The large peaks (appearing mainly for $K > 2\pi$ marked with full black circles) reflect the anomalous diffusion due to accelerator modes [of period 1, being located inside the intervals predicted by the Eq.~(\ref{acmdint})]. However, there is a number of relatively smaller peaks for $K < 2\pi$ (more clearly presented in the inset panel of Fig.~\ref{figKvsmu}), whose origin is accelerator modes of higher period as we will see later, and also for $2\pi < K < 4\pi$, both these sets are marked with empty circle. With the symbol ($\times$) we mark few typical examples, close to those peaks, for which the diffusion is normal and are also studied in detail in this section.

All the large peaks for $K > 2\pi$, marked with full black circles in Fig.~\ref{figKvsmu}, correspond to regimes with accelerator modes of period 1 and they decrease monotonically as a power law
\be
f(x)=a x^b,
\ee
where $a=2.41645$ and $b=-0.195896$ [blue dotted line in Fig.~\ref{figKvsmu}, with asymptotic standard error $\pm$0.04294 (1.777$\%$) and $\pm$0.00537 (2.741$\%$) respectively] indicating that for $K>70$ their effect decreases significantly. On the other hand, the size of the successive accelerator modes of period 1 intervals decays with a power law defined simply and analytically by the Eq.~(\ref{acmdint}).
In order to understand the effect of the presence of accelerator modes in the diffusion and transport properties of the phase space in the standard map, we first picked an, as much as possible, representative sample of $K$-values. In more detail, we included in our test-cases all the $K$-values which correspond to all the main peaks appearing in Fig.~\ref{figKvsmu} with $K > 2\pi$ together with a few cases from the `plateaus' of this curve. Furthermore, we took into account the peaks occurring for $1 \lesssim K \lesssim 2\pi$ (see the empty black circles in the inset zoom in Fig.~\ref{figKvsmu}) which are associated with accelerator modes of higher periodicity, as it will be seen thereupon. The case with $K=3.8$, whose $\mu$ value is $\approx 1$, is chosen for comparison reasons from the plateau and as it turns out has no accelerator modes in its phase space causing anomalous diffusion. Here we should stress that we repeated the same procedure for larger number of iterations $n=10000$ and it turns out that the exponent $\mu$ has well converged to the values shown in Fig.~\ref{figKvsmu}.

For each one of these $K$-values of the nonlinearity kick parameter, we performed a thorough study by calculating and comparing the following quantities

(a) The index of stability $\alpha$-parameter of the L\'evy stable distribution.\

(b) The diffusion exponent $\mu$ as described in Sec.~\ref{intro} [Eq.~(\ref{varp})].\

In principle, and, in the case of normal diffusion (Gaussian statistics) for the above quantities, one expects to find $\alpha=2$ for the L\'evy stable distribution and diffusion exponent $\mu=1$.

For the sake of completeness, one could also consider to study the correlation functions characterizing the standard map. In the literature we found statements about the inverse power law decay \cite{Ishi_etal1991PTP}, and even exponential decay (e.g. \cite{GreKau1981PRA}), but none of them is confirmed. To be more precise, we have defined the autocorrelation function of the momentum increment $u(t)=P_{t+1} - P_t = K \sin\theta_{t+1}$ as follows
\be
C(\tau) = \langle u(t+\tau) u(t) \rangle_t,
\ee
where the time averaging is over a (sufficiently long) orbit emanating from an initial condition. After many careful attempts we concluded that $C(\tau)$ does not depend significantly on the initial condition, so long as it is in the chaotic sea. The variance $C(0) = \langle u^2(t)\rangle$ is equal to $K^2/2$, within the accuracy of about one percent, precisely as expected, if the $u(t) = K \sin\theta_{t+1}$ are completely chaotic and uncorrelated quantities, $\theta_{t+1}$ being uniformly distributed in the interval $[0,2\pi]$. This has been inspected and confirmed for several values of $K$, those where  chaos is the strongest (like $K=10$) and we have Gaussian diffusion properties, as well as for those $K$, where we have accelerator modes (like $K=6.5$) and anomalous superdiffusion. Moreover, $C(\tau)$ drops from its value $C(0)$ to $C(1)$ by about a factor one hundred, which means a very fast decay of correlations. Indeed, $C(\tau)$ behaves rather erratically for $\tau \ge 1$, fluctuating wildly by about three decadic orders of magnitude around a small value, although its average value (over an ensemble) decreases slowly, roughly exponentially, but with an exponent close to zero, so it is almost linear decay for times up to $\tau=5000$. This behavior is observed for practically all initial conditions, also for the averages over various ensembles of initial conditions, and also for various parameter values $K$. Therefore, it is hard or practically impossible to extract a physically meaningful and useful information from the statistical behaviour of the tails of the correlation function $C(\tau)$. This is the reason why we have not dealt with the correlation functions any further.

\begin{figure}\centering
\includegraphics[width=\columnwidth]{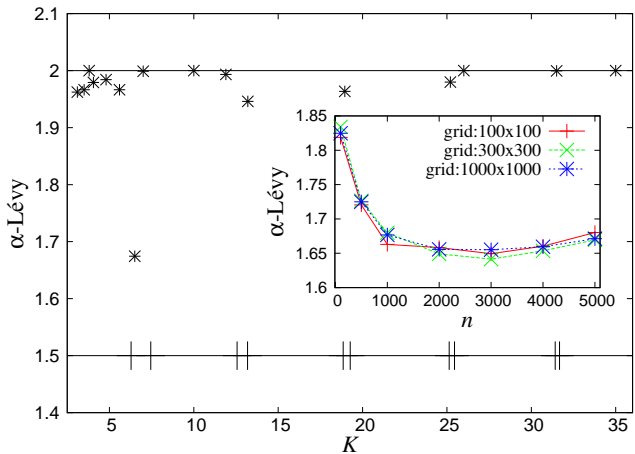}
\caption{The index of stability (or characteristic exponent) $\alpha$ of the L\'evy stable distribution for the $K$-values corresponding to cases of anomalous diffusion (peaks of the Fig.~\ref{figKvsmu}) where $\mu \ne 1$ (and generally $\alpha < 2$) and few examples where $\alpha=2$ for cases with $K$ values chosen in their vicinity (where $\mu=1$). The intervals on the black horizontal line $\alpha=1.5$ indicate the intervals of stable accelerator modes of period 1 [Eq.~(\ref{acmdint})]. Note that, for the cases where $K<2\pi$, the fit was done for an ensemble in a cell around the origin $(\theta,P)=(0,0)$ instead of a grid uniform in the entire phase space. By doing this, the data coming from islands of stability, whose momenta do not diffuse at all and mix up the distribution, are excluded. For the full set of $(\alpha,\beta,\gamma,\delta)$ fit parameters are given in Table~\ref{tb1}. In the inset we show the $\alpha$-L\'evy parameter as a function of $n$ for different cell sizes defined by the grid choices ($100\times100$, $300\times300$ and $1000\times1000$) on the plane $(\theta,P)$, for the case where $K=6.5$ and a box in the corner $(\theta,P)=(0,0)$ with $\approx$100000 ($314\times 314$) initial conditions.}
\label{figLevyK}
\end{figure}

In Fig.~\ref{figLevyK} we depict the index of stability (or characteristic exponent) $\alpha$ of the L\'evy stable distribution for the $K$-values corresponding to the cases of anomalous diffusion (peaks of the Fig.~\ref{figKvsmu}) where $\mu \ne 1$ (with $\alpha < 2$ in general) and few examples where $\alpha=2$ for cases with $K$ values chosen in their vicinity (where $\mu=1$). A special treatment was performed for the cases where $K<2\pi$, namely the fit was done for an ensemble in a cell around the origin $(\theta,P)=(0,0)$ instead of a grid uniform in the entire phase space. In this way one manages to exclude the data coming from islands of stability, whose momenta do not diffuse at all and mix up the distribution. All these values are summarized in Table~\ref{tb1}, where the whole set of the stable L\'evy distribution parameters $(\alpha,\beta,\gamma,\delta)$ is given in detail. The parameter $\alpha$ turns out to be equal to 2 for the cases ($K$-values) of phase spaces without accelerator modes. On the other hand, when such modes are present then $\alpha <2$. The $\alpha$-L\'evy parameter reaches its minimum value $\approx 1.674$ for $K=6.5$ where the effect of the accelerator mode of period 1 is the most intense, as also revealed in Fig.~\ref{figKvsmu}. In order to check whether the $\alpha$-L\'evy parameter has well converged, we tested for $K=6.5$ several different choices of the cell defined by the grid size ($100\times100$, $300\times300$ and $1000\times1000$) on which we take the initial conditions. It turns out that, by shrinking the size of each box and keeping the same number of initial conditions the $\alpha$ is practically the same (see the inset panel in Fig.~\ref{figLevyK}) after a fixed number of iterations $n=5000$.

\begin{figure*}\centering
\includegraphics[width=15cm,height=8cm]{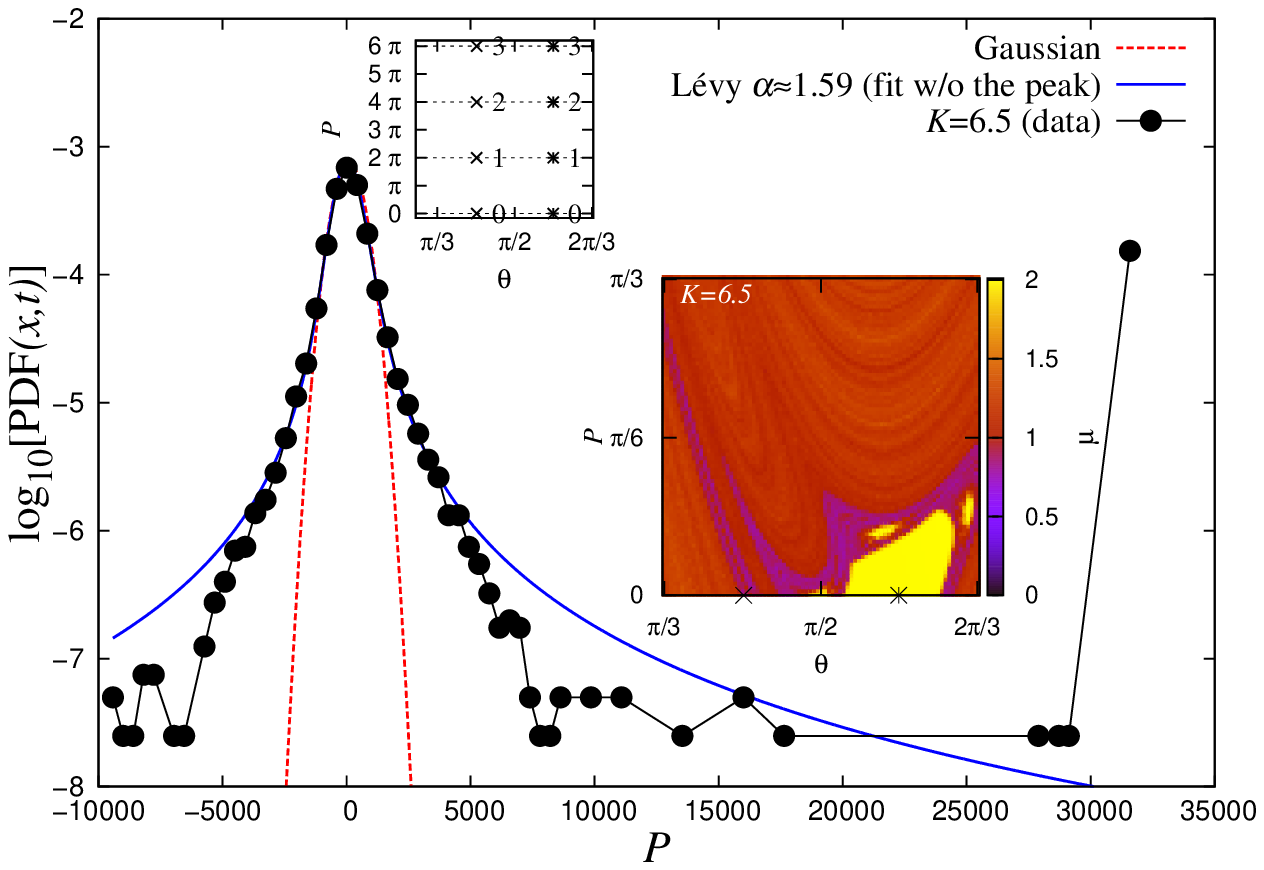}
\caption{(Color online) The L\'evy stable distribution with parameters $(\alpha,\beta,\gamma,\delta)\approx(1.59,0.164,3.7,83.63)$ for $K=6.5$ and a cell (shown in the inset) whose area contains  $\approx$100000 $(314\times 314)$ mixed initial conditions, namely trajectories that are transported ballistically by the \emph{unstable} accelerator mode around $(\theta_0,P)=(1.31179, 0)$ ($\times$) and more evidently around the \emph{stable} one at $(\theta_1,P)=(\pi-\theta_0, 0)$ ($\ast$), together with chaotic ones in their neighborhood after $n=5000$. The color-plot inset panel depicts the diffusion exponent $\mu$, calculated by the process described in Sec.~\ref{figDcln} [Eq.~(\ref{varp})], for a grid of cells on the subspace $(\theta,P) \in [\pi/3, 2\pi/3]\times[0,\pi/3]$ of the phase space. The yellow (light gray in b/w) color areas corresponds to ballistic motion due to the two accelerator modes pointed by the symbols $(\times)$ and $(\ast)$ respectively. The ensembles of initial conditions lying inside the ``pure'' chaotic regime diffuse normally with $\mu \approx 1$. The ``belt''-like darker zone in the edges of the accelerator mode area with $\mu \approx 0.8$ indicates a sticky subdiffusive transport. In the small inset panel, we show the ballistic transport on the cylindrical phase space of the two above mentioned accelerator modes positioned initially at $(\theta_{0,1},0)$ and boosted by $\delta P = 2 \pi$ at every successive kick.}
\label{figACmode}
\end{figure*}

In Fig.~\ref{figACmode} we show the stable L\'evy distribution with parameters $(\alpha,\beta,\gamma,\delta)\approx(1.59,0.164,3.7,83.63)$ for $K=6.5$ and a cell whose area contains $\approx$100000 $(314\times 314)$ mixed initial conditions, i.e. trajectories that are transported ballistically by the  \emph{unstable} accelerator mode around $(\theta_0,P)=(1.31179, 0)$ ($\times$) and more evidently around the \emph{stable} one at $(\theta_1,P)=(\pi-\theta_0, 0)$ ($\ast$) together with chaotic ones in their vicinity after $n=5000$. The fit here has been performed by excluding the last peak in the positive large momenta due to the ballistic transport by the accelerator mode. In general there can be two peaks in the distribution of the diffusive variable $P$ (in the positive and negative direction) depending on the choice of the ensemble of initial conditions. Nevertheless, as also explained thoroughly in \cite{Stef_etal1998PRE}, the distribution will become an $\alpha$-L\'evy stable one and the peaks recede to infinity with amplitudes tending to zero. The color-plot inset panel depicts the diffusion exponent $\mu$, calculated by the process described in Fig.~\ref{figDcln}, for a grid of cells on the subspace $(\theta,P) \in [\pi/3, 2\pi/3]\times[0,\pi/3]$ of the phase space. The yellow (light gray in b/w) color areas correspond to ballistic motion due to the two accelerator modes pointed by the symbols $(\times)$ and $(\ast)$ star respectively. The ensembles of initial conditions lying inside the fully chaotic regime diffuse normally with $\mu \approx 1$. Moreover, there is a ``belt''-like darker zone in the edges of the accelerator mode area with $\mu \approx 0.8$ which indicates a sticky subdiffusive transport. In the small inset panel we show the ballistic transport on the cylindrical phase space of the two above mentioned accelerator modes positioned initially at $(\theta_{0,1},0)$ and boosted by $\delta P = 2 \pi$ at every (only four here) successive kick.

\begin{figure*} \centering
\includegraphics[width=5.75cm]{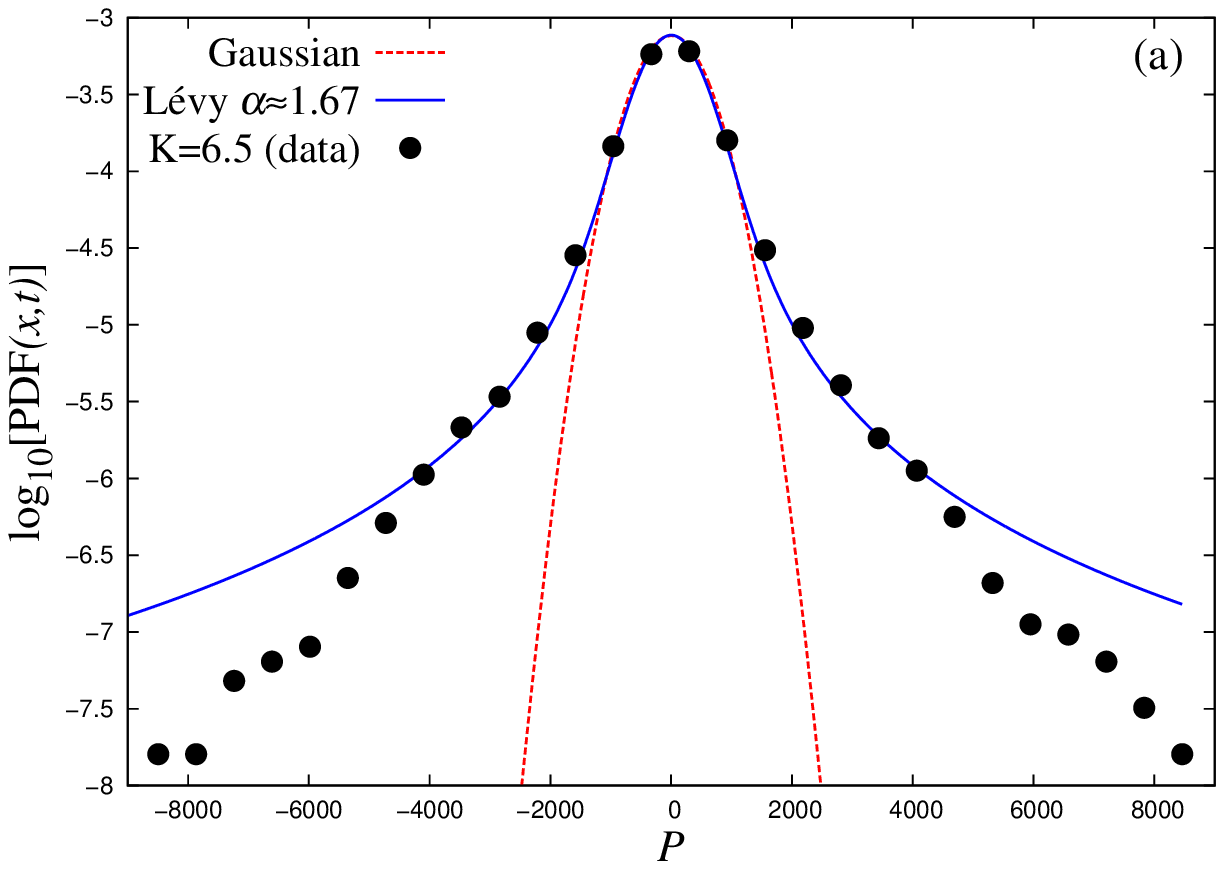}
\includegraphics[width=5.75cm]{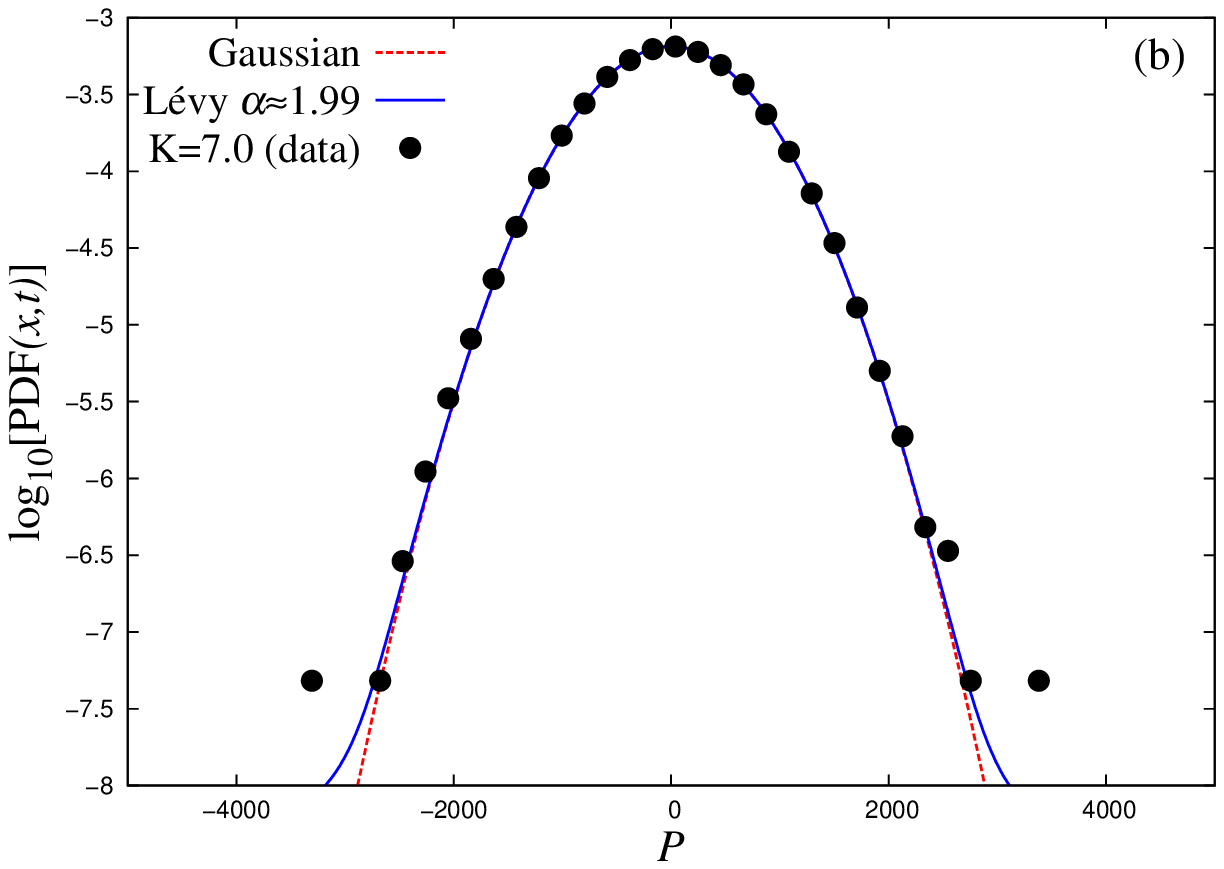}
\includegraphics[width=5.75cm]{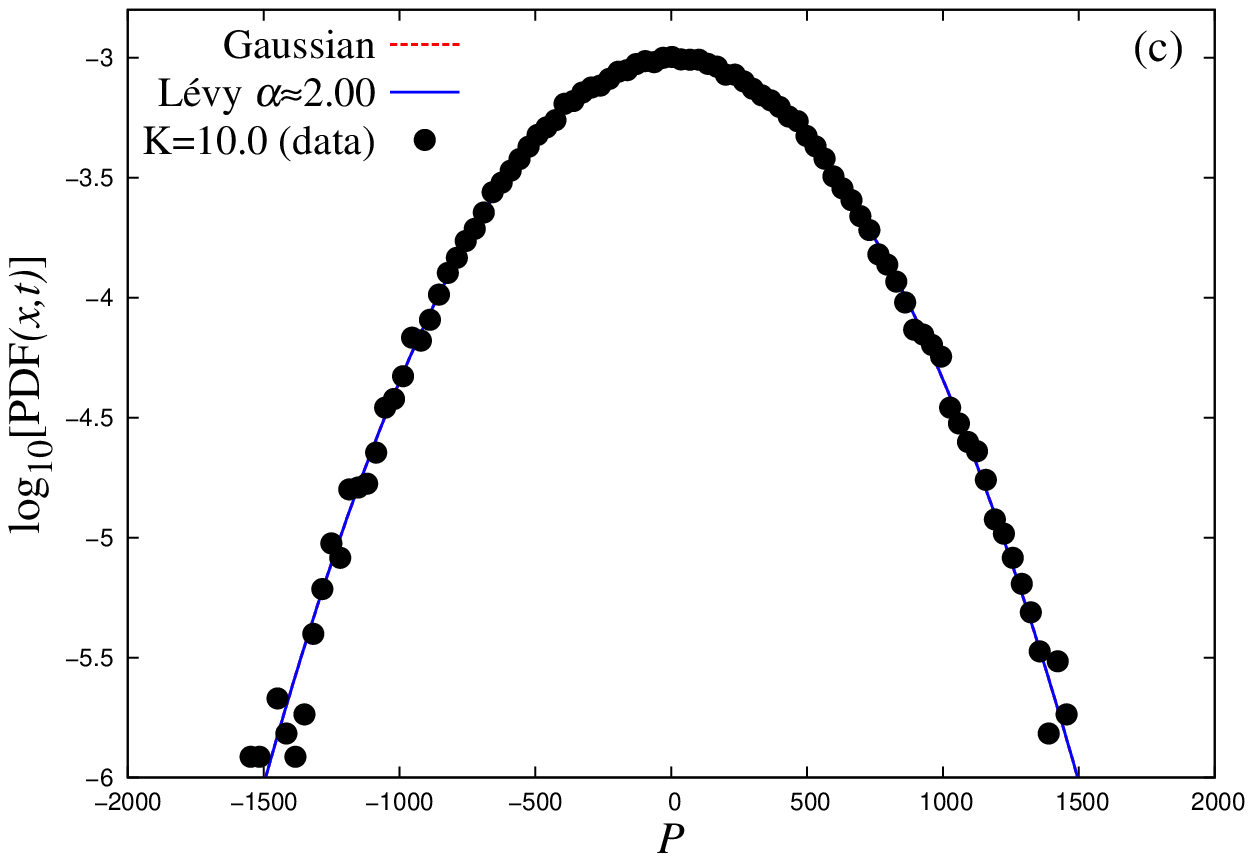}
\includegraphics[width=5.75cm]{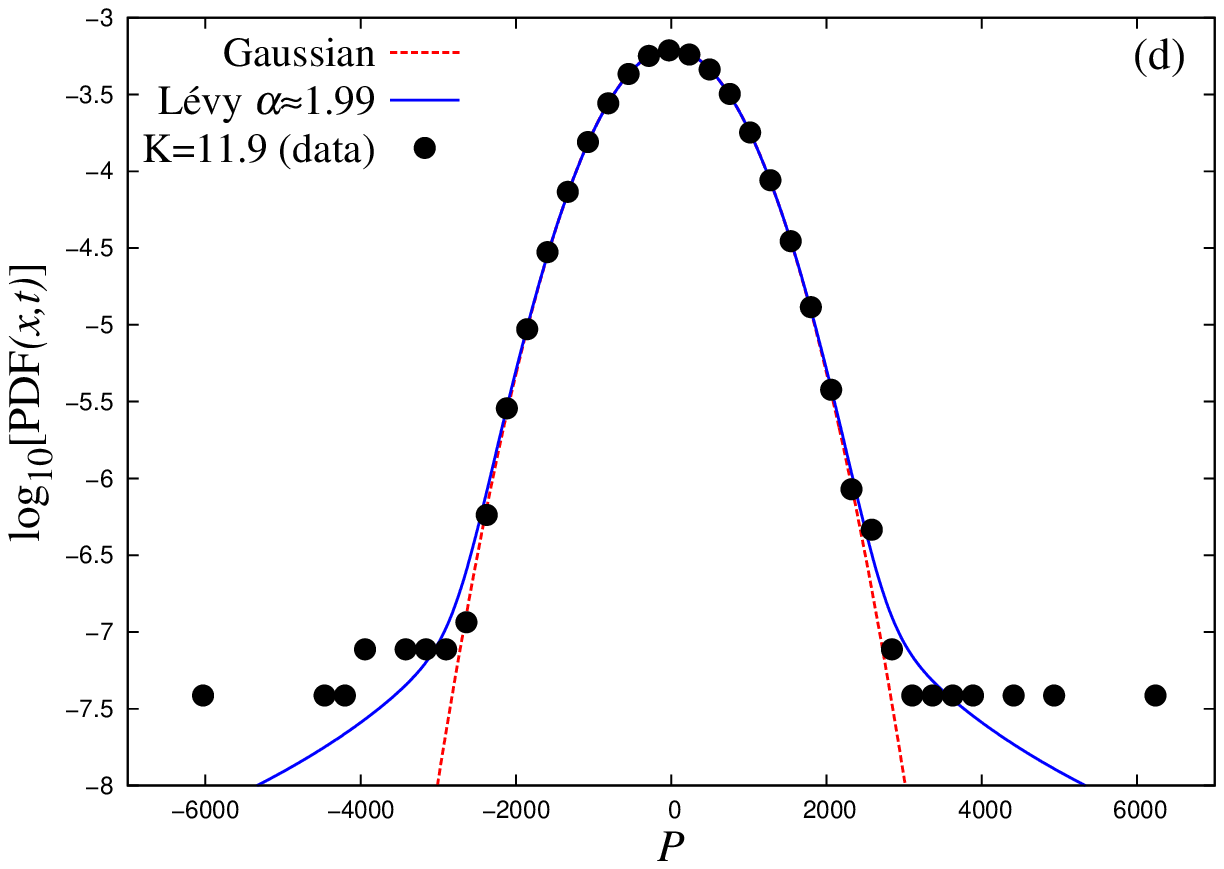}
\includegraphics[width=5.75cm]{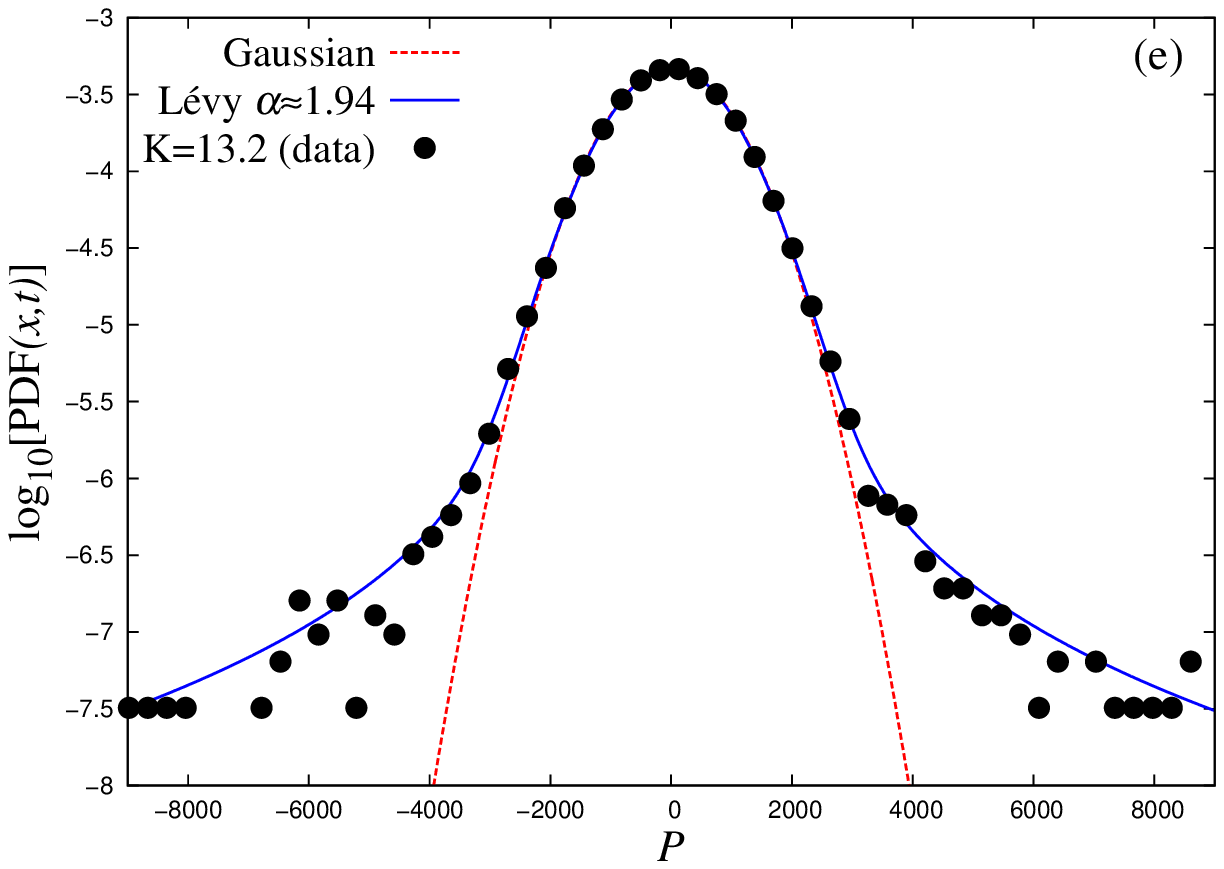}
\includegraphics[width=5.75cm]{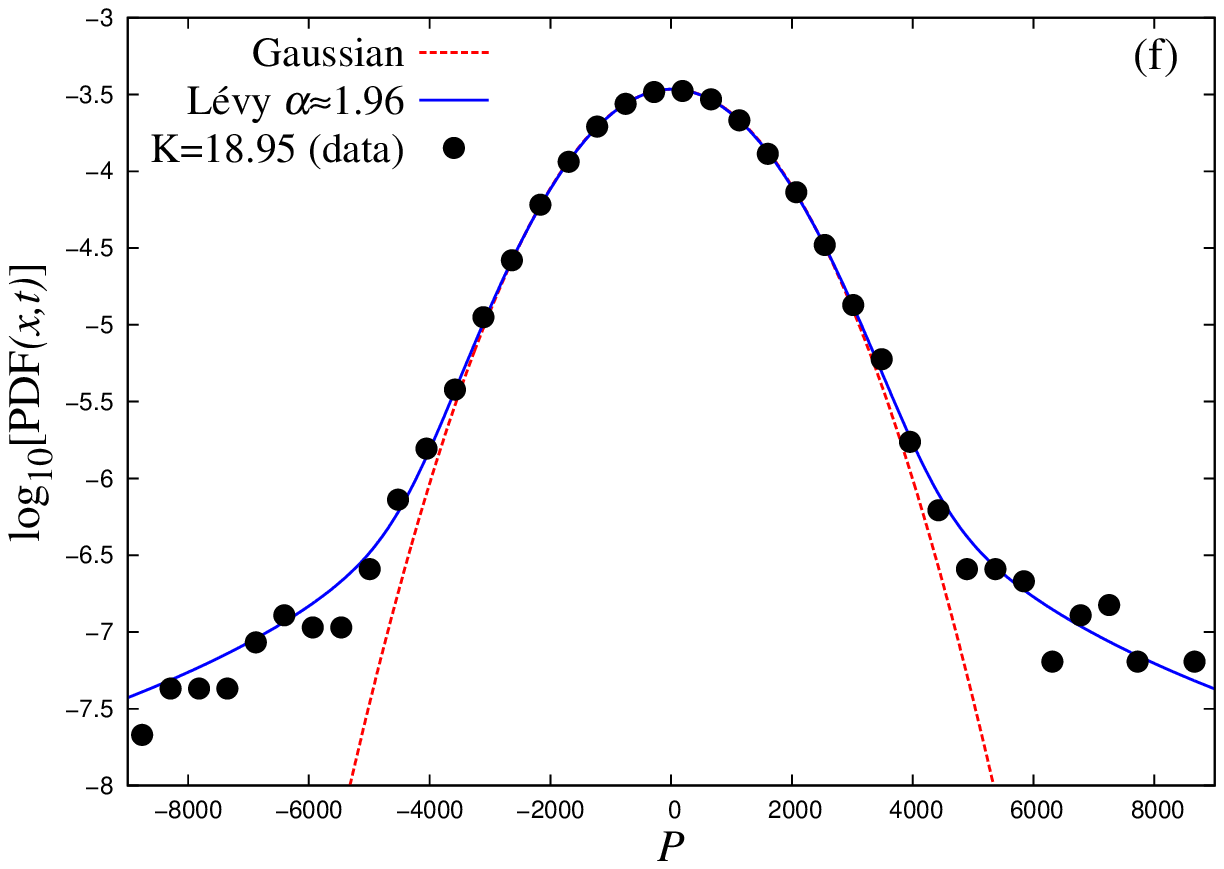}
\caption{(Color online) The probability distribution function of the momentum $P$ after $n=5000$ iterations (black filled circles) and the fits with the stable $\alpha$-L\'evy distribution for a sample of $K$-values associated with the principal peaks (presence of accelerator modes of period 1) of the Fig.~\ref{figKvsmu}, i.e., $K=6.5,11.9,13.2,18.95$ (a),(d),(e),(f) where $\alpha$ is generally not equal to 2 and two $K$-values without accelerator modes, i.e., $K=7.0,10.0$ (b),(c) where $\alpha$ is equal to 2. Here, the Gaussian distribution (red dashed line) for comparison reasons with the best-fit function, depicted in the figures, is derived by the $S(\alpha,\beta,\gamma,\delta;0)$ probability distribution by setting $\alpha=2$ and keeping all the rest parameters same as the given by the fits in general $\alpha \ne 2$. The total number of initial conditions is $\approx$100000 ($314 \times 314$) on a uniform grid in the entire phase plane $(\theta,P) \in [0,2\pi] \times [0,2\pi]$. For larger $K$-values with accelerator modes of period 1, their effect in the diffusion process is becoming gradually weaker, as can be seen also by the decay of the $\mu$ value in Fig.~\ref{figKvsmu}, and the probability distributions tend to Gaussian-like ones.}
\label{figLevyHisto}
\end{figure*}

In Fig.~\ref{figLevyHisto} we present the probability distribution function of the momentum $P$ after $n=5000$ iterations (black filled circles) and the fits with the $\alpha$-L\'evy stable distribution (solid blue line) for a sample of $K$-values associated with the principal peaks (presence of accelerator modes of period 1) of the Fig.~\ref{figKvsmu}, i.e., $K=6.50,11.90,13.20,18.95$. In Fig.~\ref{figLevyHisto}(a),(d),(e),(f) $\alpha$ is generally not equal to 2 and two $K$-values without accelerator modes, i.e., $K=7.0,10.0$, are shown in Fig.~\ref{figLevyHisto}(b),(c) where $\alpha$ is equal to 2. Here, and for comparison reasons with the best-fit function depicted in the figures, the Gaussian distribution (red dashed line) is derived by the $S(\alpha,\beta,\gamma,\delta;0)$ probability distribution by setting $\alpha=2$ and keeping all the remaining parameters the same as given by the fits generally for $\alpha \ne 2$. The total number of initial conditions is $\approx$100000 $(314 \times 314$) on a uniform grid in the entire phase plane $(\theta,P) \in [0,2\pi] \times [0,2\pi]$. For larger $K$-values with accelerator modes of period 1, their effect in the diffusion process is becoming gradually weaker, as can also be seen by the decay of the $\mu$ value in Fig.~\ref{figKvsmu}, and the probability distributions tend to Gaussian-like ones. We have also calculated the $\chi^2$-test for all the density distributions of Fig.~\ref{figLevyHisto}. However, the derived quantitative information turns out, not to differ much to what the panels show. Let us also stress that the goodness of the several fits is not the main point here but rather the deviation from the Gaussian profile which is evident when the effect of the accelerator mode is relatively strong as shown in panel (a) and expressed by $\alpha$ parameter value far from 2. Moreover, due to our main motivation, i.e., to focus on the relevant quantum time scales, we do not expect to capture very accurately the tails, which are expected to follow the theoretical curve at much larger times. More examples and the details regarding the whole set of parameters can be found in Table~\ref{tb1}. Note that, for $K<2\pi$ the fit was done for an ensemble in a cell around the origin $(\theta,P)=(0,0)$ instead of a grid uniform in the entire phase space, in order to exclude the data coming from islands of stability whose momenta do not diffuse at all and mix up the distribution.
\begin{table}\caption{\label{tb1}
L\'evy stable distribution parameters for various values of $K$ up to $K=35$ (indicated with full, empty circles and crosses in Fig.~\ref{figKvsmu}). Note that, for the L\'evy stable distribution parameters with $K<2\pi$ the fit was done for an ensemble in a cell around the origin $(\theta,P)=(0,0)$ instead of a grid uniform in the entire phase space, in order to exclude the data coming from islands of stability whose momenta do not diffuse at all and mix up the distribution.}
\begin{ruledtabular}
\begin{tabular}{c|cccc}
\multicolumn{5}{c} {\rm \ \ \ \ \ \ L\'evy stable distribution} \\
$K$ & $\alpha$ & $\beta$ & $\gamma$ & $\delta$ \\
\colrule
\ 3.10 & $\approx$ 1.962 &  $-$0.001 &  98.0 & \ 1.08 \\
\ 3.50 & $\approx$ 1.966 & \ 0.083 &  99.0 & \ 3.79  \\
\ 3.80 & $\approx$ 2.00 & \ 0.000 & 107.3 & \ 6.10 \\
\ 4.05 & $\approx$ 1.979 &  -0.075 & 132.6 & \ 6.12 \\
\ 4.80 & $\approx$ 1.984 & \ 0.039 & 170.7 & \ 4.98 \\
\ 5.60 & $\approx$ 1.966 & \ 0.0577 & 234.5 & \ 6.76 \\
\ 6.50 & $\approx$ 1.674 & \ 0.000 & 369.2 & \ 1.07 \\
\ 7.00 & $\approx$ 1.998 &  -0.084 & 433.3 & \ 0.63 \\
10.00 & $\approx$ 2.000 & \ 0.000 & 284.4 & \ 1.56 \\
11.90 & $\approx$ 1.993 &  0.000 & 453.5 & \ 2.04 \\
13.20 & $\approx$ 1.946 &  -0.011 & 600.1 & \ 3.15 \\
18.95 & $\approx$ 1.963 & \ 0.062 & 824.7 &  10.19 \\
25.20 & $\approx$ 1.979 &  -0.031 & 1038.8 & \ 3.09 \\
26.00 & $\approx$ 2.000 & \ 0.000 & 1077.6 & \ 3.41 \\
31.50 & $\approx$ 1.999 & \ 0.000 & 1290.9 & \ 3.34 \\
35.00 & $\approx$ 2.000 & \ 0.000 & 1085.0 &  16.13 \\
\end{tabular}
\end{ruledtabular}
\end{table}

\begin{figure*} \centering
\includegraphics[width=6.cm]{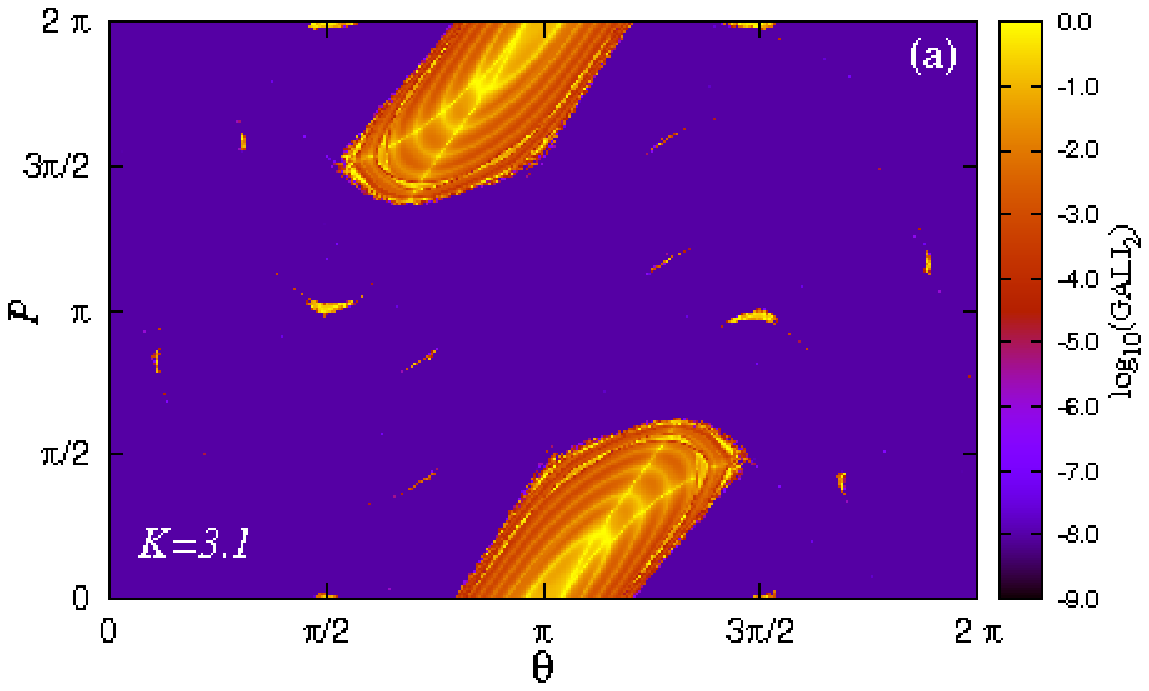}
\includegraphics[width=6.cm]{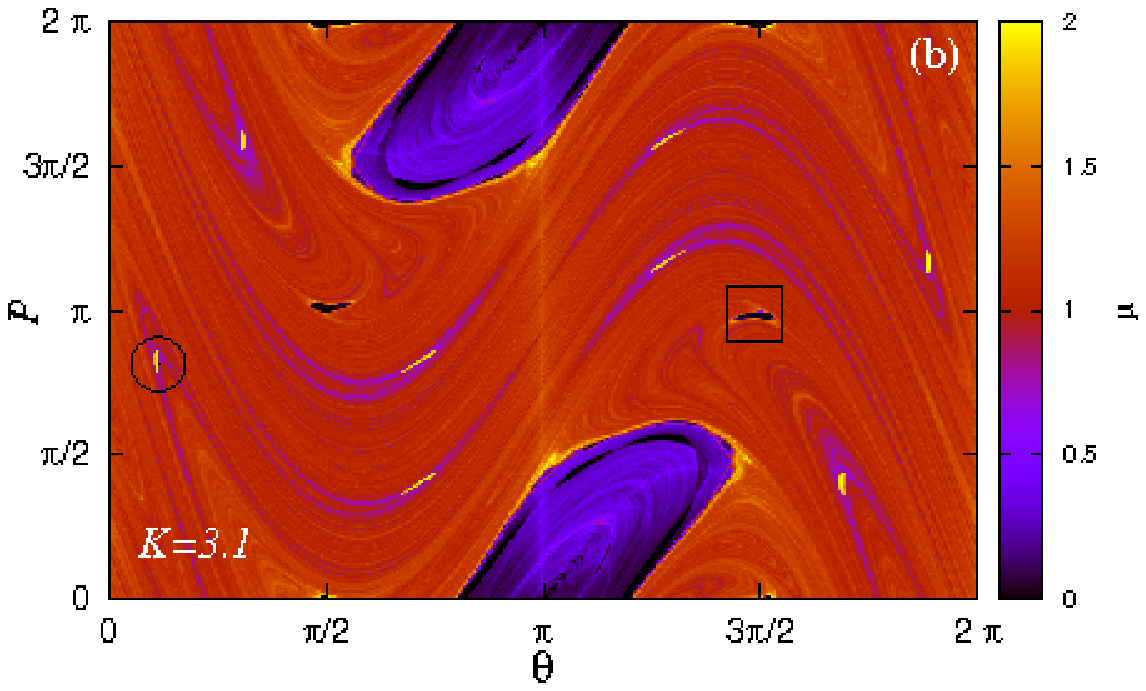}
\includegraphics[width=5.15cm,]{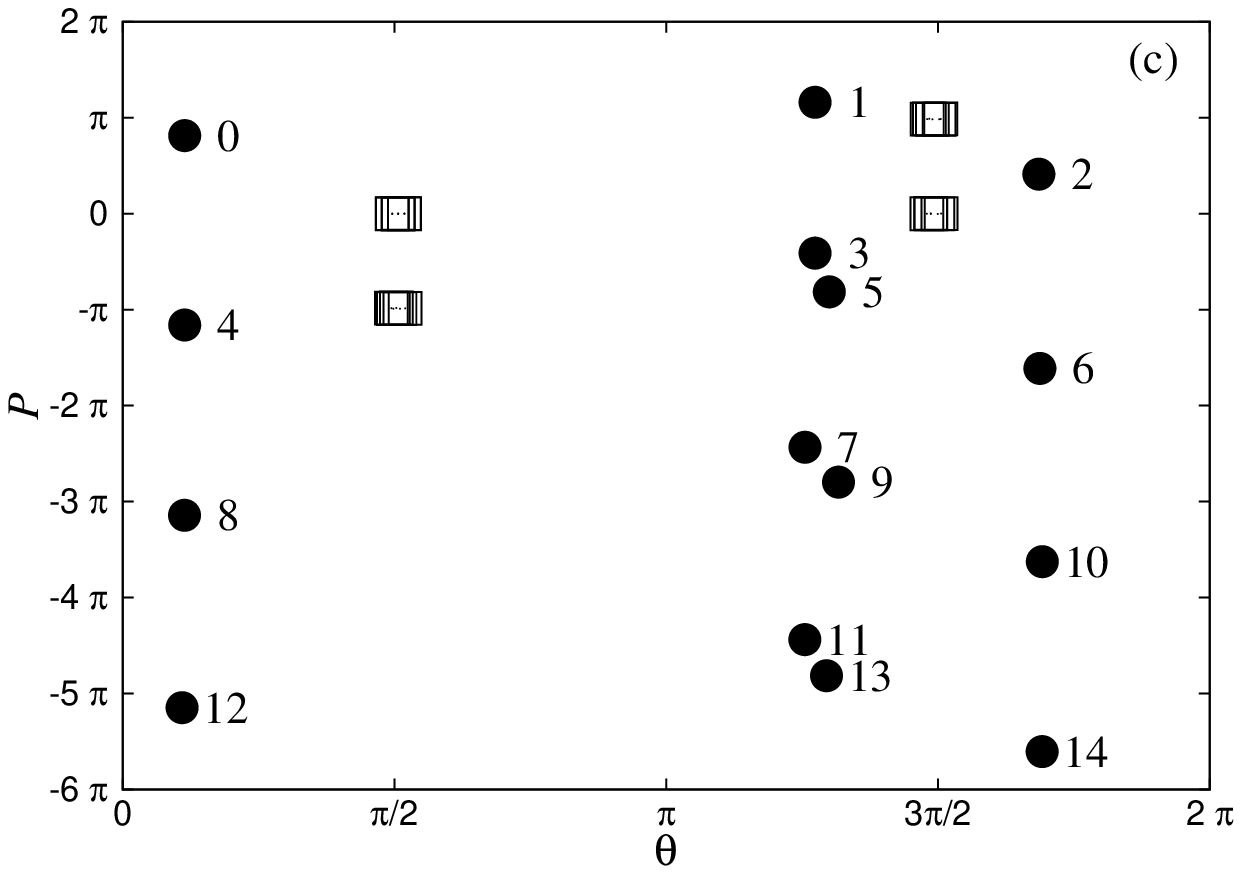}
\caption{(Color online) (a) The GALI$_2$ for $K=3.1$ with $50\times50$ initial conditions on a grid $500\times 500$ on the entire phase space $(\theta,P) \in [0,2\pi] \times [0,2\pi]$. (b) the diffusion exponent $\mu$ for the same kick parameter value for $50 \times 50$ initial conditions on a $500\times 500$ cell grid of the entire phase space calculated after $n=5000$ iterations (see text for more details). (c) Two examples (marked with an empty square and empty circle in panel b) following different diffusion processes: a trajectory transported ballistically (with $P<0$) by the effect of an accelerator mode of period 4 ($\bullet$) and one oscillating between islands of stability ($\boxdot$) of period 4 too.}
\label{figGDK31}
\end{figure*}

In Fig.~\ref{figGDK31}(a) we present the outcome of the calculation of the GALI$_2$ on the whole phase plane $(\theta,P) \in [0,2\pi]\times [0,2\pi]$ for $K=3.1$, for $500\times 500$ initial conditions uniformly distributed. Each initial condition is colored according to the color scale seen at the right side of the panel. For chaotic orbits, having small GALI$_2$ value ($\approx 10^{-8}$) are colored black, while the yellow (light gray in b/w) color corresponds to regular motion, found here to be $\approx 13.52\%$ of the whole plane, with high - close to zero - values (the color bar is in a logarithmic scale). Thus, we can clearly identify even tiny regions of regular motion which are not easily seen in the phase space portraits given often by simple Poincar\'e surface of sections. Having located the stable region of the phase space, the next point of interest is to distinguish among them, those that are due to accelerator modes from those due to islands of stability.

The distinction can be efficiently achieved with the use of the diffusion exponent $\mu$ color-plot in Fig.~\ref{figGDK31}(b), where we first consider again a grid of $500\times 500$ cells (on the entire phase space $(\theta,P) \in [0,2\pi]\times [0,2\pi]$) with $50 \times 50$ initial conditions in each and evolve all of them together for $n=5000$ iterations. Then, for each ensemble of each cell separately, we calculate numerically the diffusion coefficient $D_{\mu}(K)$ as a function of the iterations $n$ and perform a fit procedure just like in Fig.~\ref{figDcln} to calculate the diffusion exponent $\mu$ [represented in the color bar of Fig.~\ref{figGDK31}(b)] that \emph{characterizes} the different kind of diffusion of this small area. Depending on the relative location of each ensemble one may expect to find: (i) \textit{ normal diffusion} ($\mu = 1$) inside chaotic regimes without the presence of accelerator modes, (ii) \textit{subdiffusion} ($0< \mu < 1$) inside islands of stability, (iii) \textit{superdiffusion} ($1 < \mu < 2$) inside chaotic regimes with the presence of accelerator modes in the phase space and (iv) \textit{ballistic transport} ($\mu \approx 2$) inside and in the very close vicinity of accelerator modes.

It turns out that the stable regions around $(\pi/2,0)$, $(3\pi/2,0)$, $(\pi/2,\pi)$, $(3\pi/2,\pi)$ and $(\pi/2,2\pi)$, $(3\pi/2,2\pi)$ are indeed islands of stability since their diffusion exponent $\mu$ is smaller than 1. The remaining tiny stable areas correspond to stable higher period accelerator modes with $\mu \approx 2$. Here we manage to locate the accelerator modes of higher periods 2,3,4,... which in general are not so easy to be calculated analytically. In Fig.~\ref{figGDK31}(c) we show two examples (marked with an empty square and empty circle in panel b) following different diffusion processes, i.e., for a trajectory oscillating between islands of stability (empty dotted boxes) and for one transported ballistically (with $P<0$) by the effect of an accelerator mode (full circles). The period of both is 4 (when projected on the $(\theta,P) \in [0,2\pi]$), as it can be seen by iterating them for a few steps.

\begin{figure*}\centering
\includegraphics[width=8.5cm]{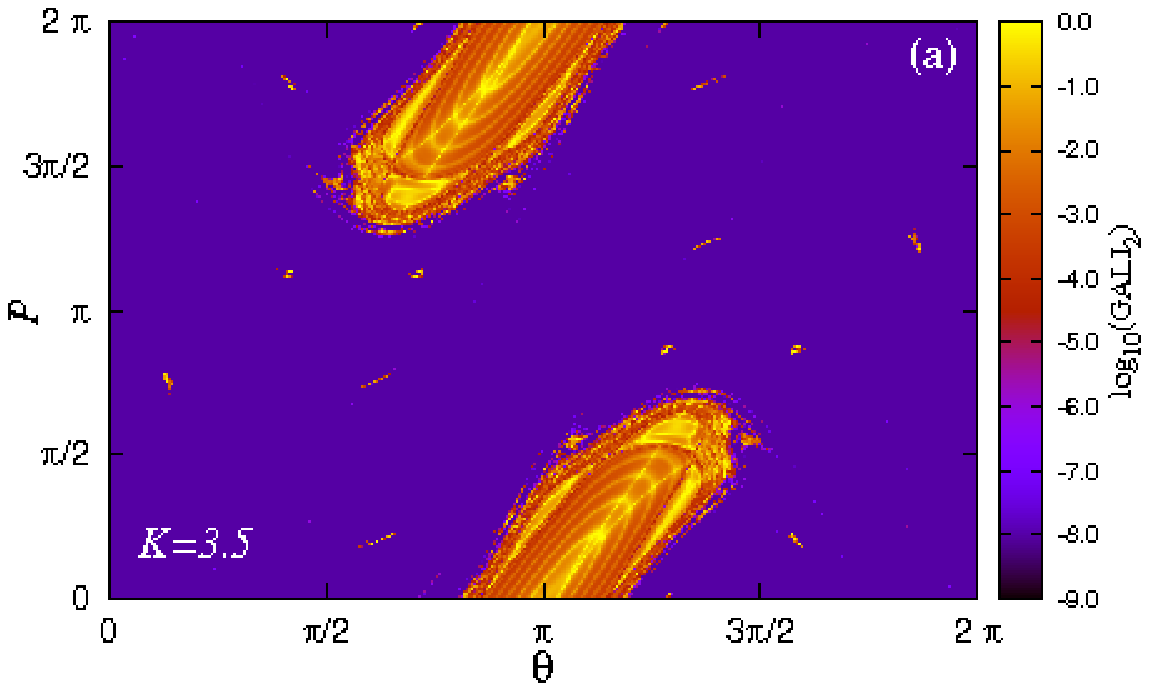}
\includegraphics[width=8.5cm]{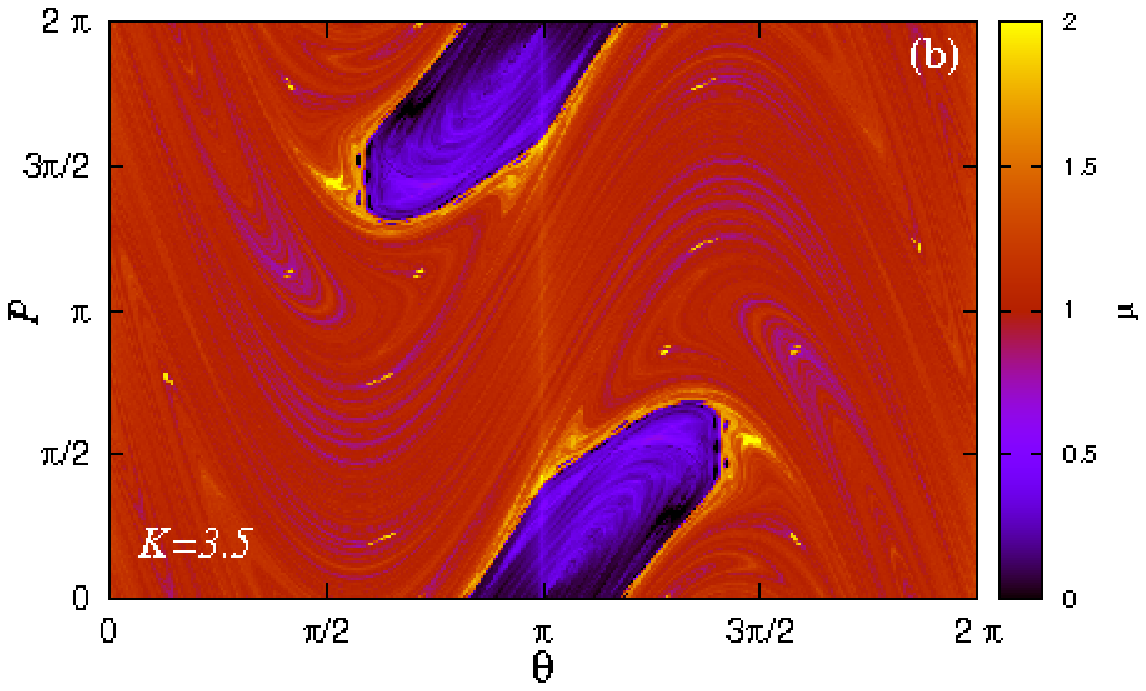}
\includegraphics[width=8.5cm]{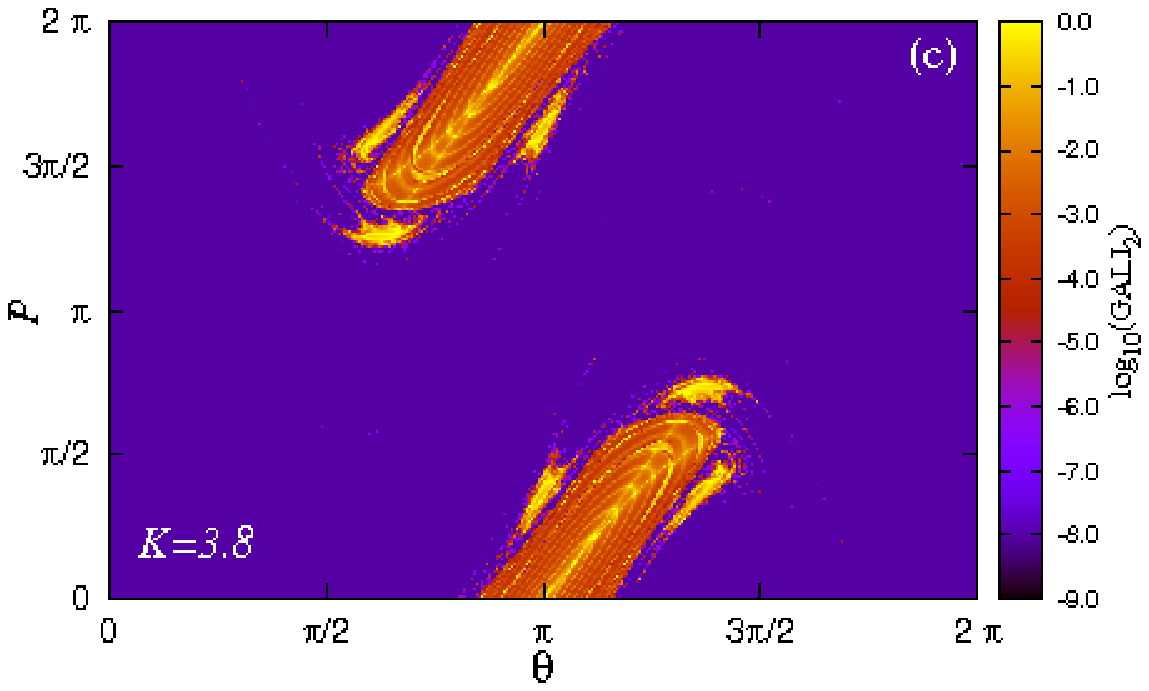}
\includegraphics[width=8.5cm]{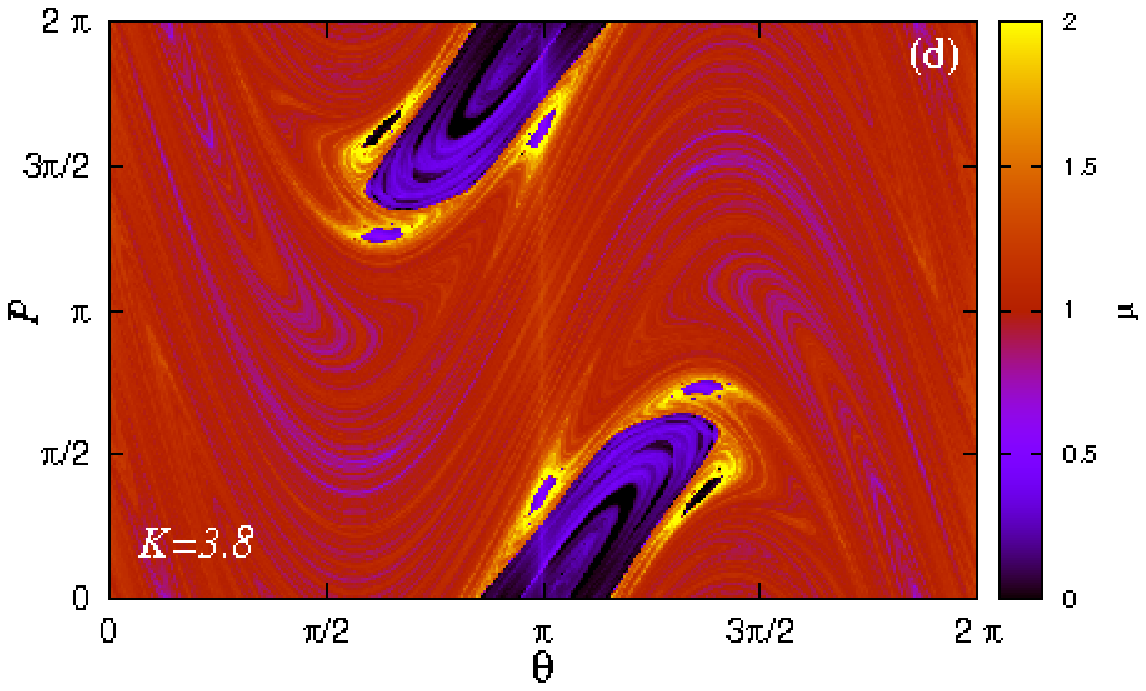}
\includegraphics[width=8.5cm]{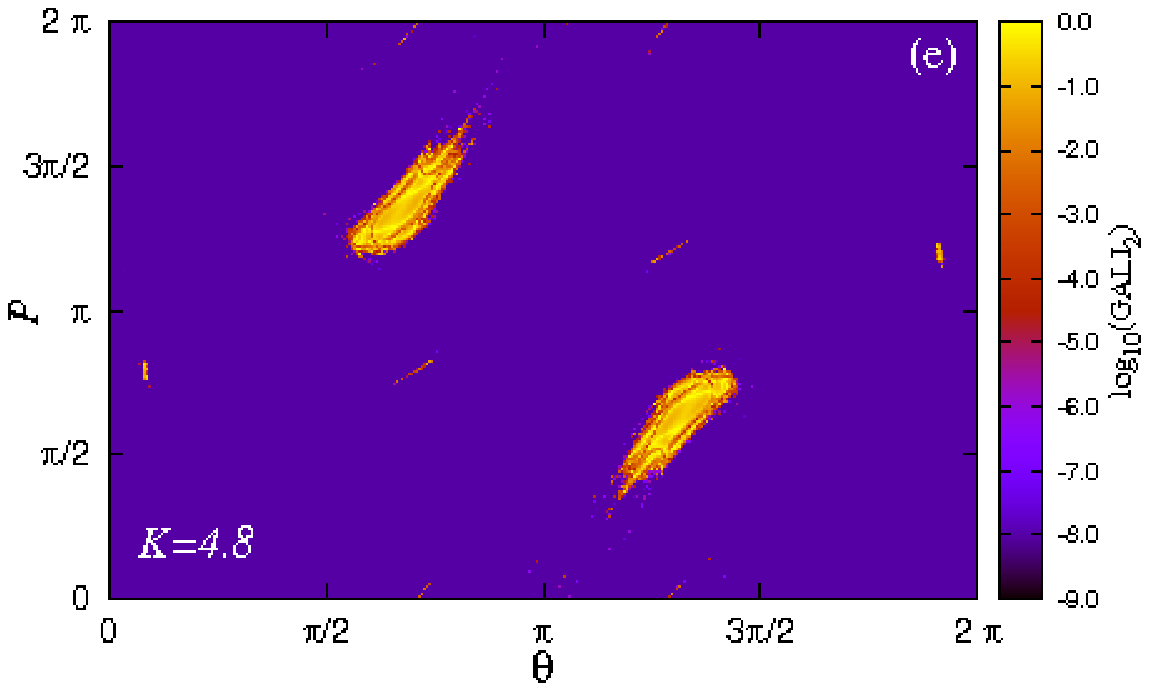}
\includegraphics[width=8.5cm]{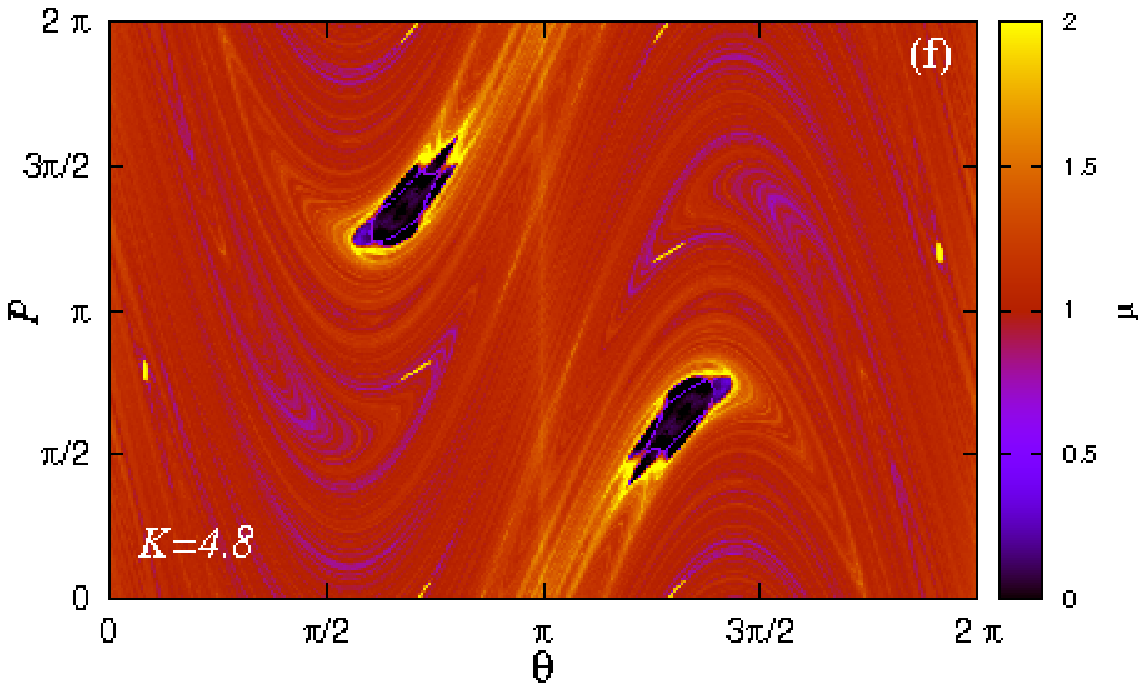}
\includegraphics[width=8.5cm]{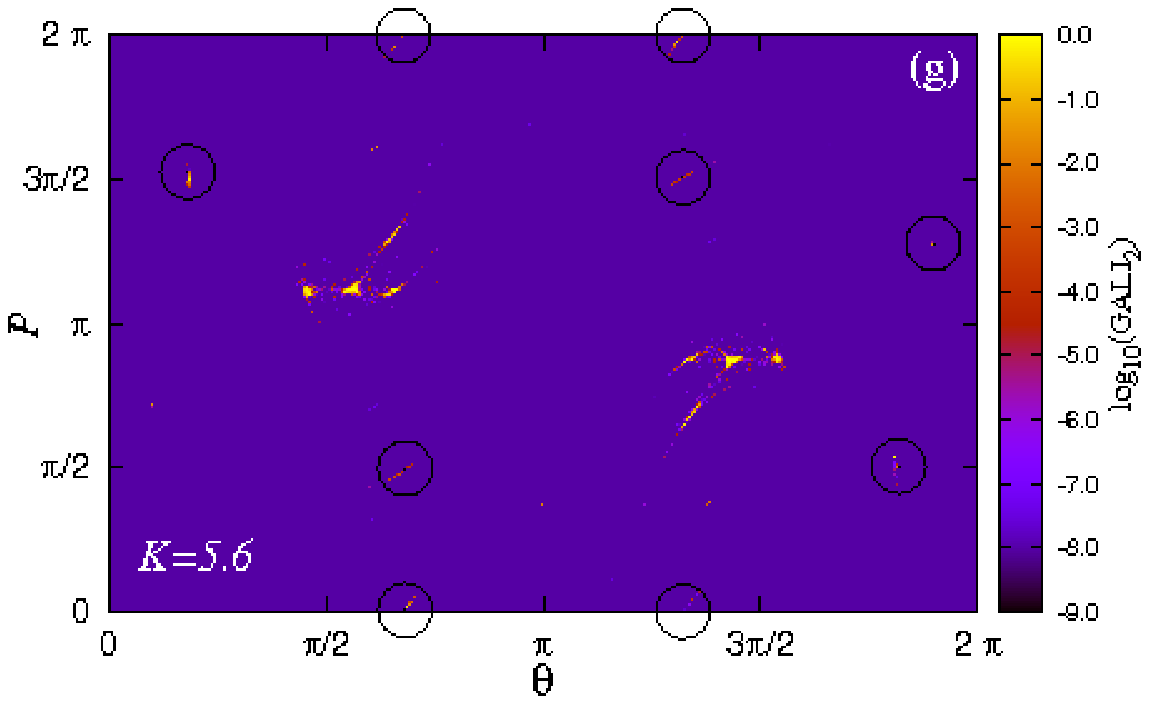}
\includegraphics[width=8.5cm]{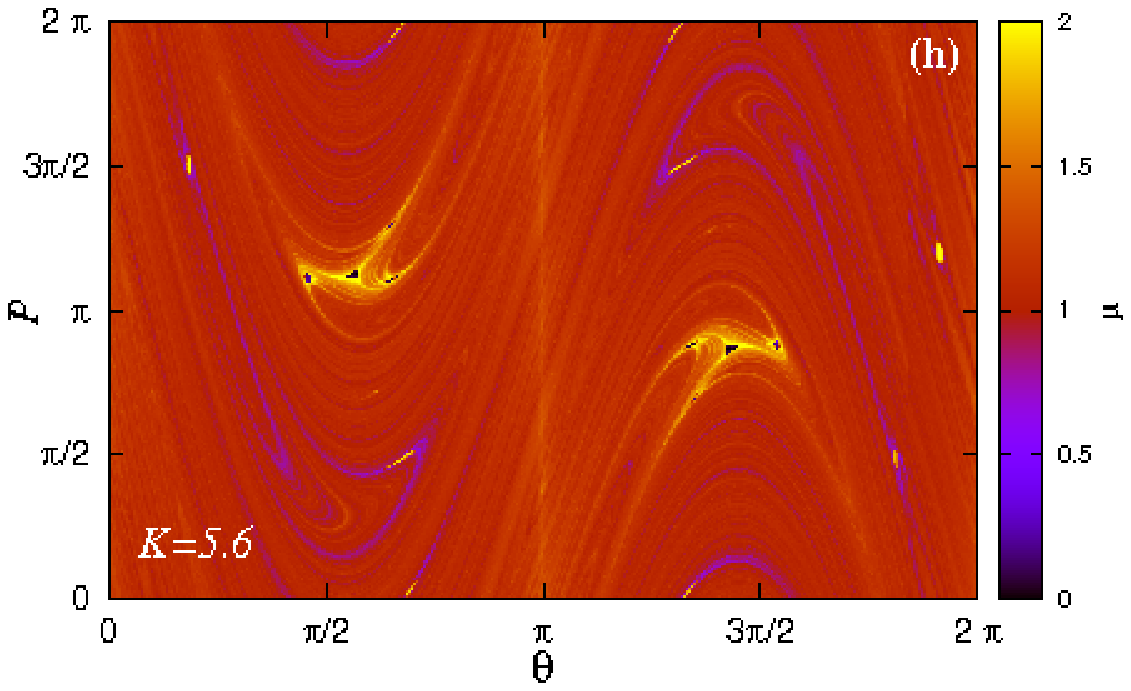}
\caption{(Color online) Same as \ref{figGDK31} for  $K=3.5$ (1st row), $K=3.8$ (2nd row), $K=4.8$ (3rd row), $K=5.6$ (4th row). The yellow (light gray in b/w version) areas scattered in the large chaotic sea \textit{in both} panels of each row correspond to accelerator modes of higher period, while those being yellow (light gray in b/w) in the GALI color-plot \textit{and} at the same time dark blue or black in the diffusion exponent $\mu$ color-plot, are islands of stability. There are also orbits in the edges of the big islands of stability which are transported non-normally along the cylindrical phase space. Note that for $K=3.8$ (2nd row) is chosen as a counter example for comparison from the interval of the kick parameter values where the diffusion for ensembles chosen inside the chaotic sea is normal $\mu \approx 1$ without the presence of accelerator modes. For the case of $K=5.6$ (4th row) the two relatively large stable areas located in their centers correspond to small non-diffusive islands of stability while those tiny and rather hardly visible (marked with circles in panel g) to ballistic accelerator modes (yellow or light gray in b/w in panel h).}
\label{figGDKs}
\end{figure*}

In Fig.~\ref{figGDKs} we show a few more examples for $K=3.5$ (1st row), $K=3.8$ (2nd row), $K=4.8$ (3rd row), $K=5.6$ (4th row), charting the phase space's stable and chaotic regions with the GALI$_2$, and the normal or anomalous diffusion with the exponent $\mu$ in a similar manner as in Fig.~\ref{figGDK31}. The yellow (light gray in b/w) areas scattered in the large chaotic sea \textit{in both} panels of each row correspond to accelerator modes, while those being yellow (light gray in b/w) in the GALI color-plot \textit{and} at the same time dark blue or black in the diffusion exponent $\mu$ color-plot, are islands of stability. There are also orbits in the edges of the big islands of stability which are transported non-normally along the cylindrical phase space, due to the accelerator modes of higher period.

For the case with $K=3.5$ (1st row) and in panel (a), we show all the stable (colored yellow or light gray in b/w) areas as detected by the GALI method, whose total relative fraction in the whole phase space is $\approx 14.12\%$. When comparing to the diffusion exponent color-plot of panel (b) we can furthermore see that, all the small scattered stable regions of panel (a) are due to higher period accelerator modes having $\mu \approx 2$. The case with $K=3.8$ (2nd row) is a counter example for comparison from the interval of the kick parameter values where the diffusion for ensembles chosen inside the chaotic sea is normal $\mu \approx 1$ and $\alpha$-L\'evy $\approx 2$, since no accelerator modes are present. The relative fraction of regular motion here is estimated to be $\approx 12.29\%$. Obviously, there are some accelerator modes (colored in yellow or light-gray in b/w) of higher period around the three small islands of stability in the upper and lower part of the figure. However, they affect only locally the diffusion, in the sense that the vast phase space does not ``feel'' their presence. Regarding the case with $K=4.8$ (3rd row), we see that the relative size of stable motion has become even smaller (two dominant islands of stability in the central part with $\approx 2.93\%$ of the phase space) while only a few small scattered stable regions in the large chaotic see have remained compared to the case of $K=3.8$ (and $K=3.1$ [Fig.~\ref{figGDK31}(a),(b)]). For $K=5.6$ (4th row) the stable islands in the phase space have shrunk even more (see panel g) with the relative fraction of regular motion to be now only $\approx 0.62\%$. Combining the panels (g) and (h), it turns out that the two relatively large stable areas located in their centers correspond to small non-diffusive islands of stability while those tiny and rather hardly visible (marked with circles in panel g) to ballistic accelerator modes (yellow or light gray in b/w in panel h). In conclusion, we see that close to the boundary but outside the large island of stability, there might be both the accelerator modes of higher period surrounded by tiny second-order stable regions, and regions of stickiness which originates from the cantori. Which of the two structures prevails in the diffusion process is a very intricate question, which should be studied further in a future work.

Regarding the diffusion exponent color-plot for the cases of $K$ where accelerator modes are present in the phase space, one can notice certain (purple) spiral features (mainly located around them) whose ensembles of initial conditions appear to be subdiffusive with $\mu \approx$ 0.8 or 0.9. Their shape is originated by a transient sticky transport process on a cylindrical phase space which takes place around these modes. By increasing the final number of iterations sufficiently enough, the $\mu$ for such ensembles tends to 1 due to the mixing process between these sticky objects and those lying in the large chaotic sea, where  normal diffusion is present. However, as discussed already in the introduction, in this paper we are particularly interested in classical time scales relevant for the quantum ones.

\section{Discussion and conclusions  \label{sec:conc} }

The main goal of the present work is to explore systematically all the  dynamical and statistical aspects of the {\em generalized diffusion} in the standard map of Chirikov (which is the Poincar\'e map of the classical kicked rotator), as a paradigm of other area preserving maps, using the various computational methods to characterize the most important features in the phase space and in the parameter space. In doing this we have particularly analyzed the role of the accelerator modes, of period one and of higher periods, in the phase space cylinder, in which the (rotation) angle $\theta$ is in the interval $[0,2\pi]$, thus measured always modulo $2\pi$, whilst the (angular) momentum
$P$ is unlimited in $(-\infty,+\infty)$.

In our case of the standard map the control parameter is the kick parameter $K$. For $K<K_c \simeq 0.9716$ there is no global transport and no diffusion in the phase cylinder, because there exist the KAM invariant (spanning) curves acting as absolute barriers. Nevertheless, the local diffusion is possible on the chaotic components around and also within the islands of stability. In such local diffusion picture it turns out that the diffusion is generally not normal, but typically subdiffusion, with the diffusion exponent $\mu < 1$, due to the sticky objects, mainly cantori, surrounding the islands of stability.

When $K>K_c$ the last global invariant curve is destroyed and the global transport and diffusion becomes possible. Due to the great complexity of the phase space, the diffusion is largely anomalous, both, either superdiffusion with $\mu >1$ or subdiffusion with $\mu<1$. The subdiffusion is due to the
sticky objects already mentioned. However, the superdiffusion is governed by the same stickiness, but now around the regular islands which are accelerator modes, that is, they increase momentum $P$ by $2\pi$ or integer multiple of
$2\pi$, in a finite number of iterations (length of the period). For $K_c \le K \le 2\pi$ these are accelerator modes of higher periodicity, whilst for $K>2\pi$ they are almost exclusively the period one accelerator modes (there are just two notable exceptions, as seen in Fig.~\ref{figKvsmu}). The stickiness due to the cantori and the tiny stability regions supporting the accelerator modes of higher periodicity can indeed coexist in the area close  but outside to the edge of the large stable islands, as is clearly seen in Fig.~\ref{figGDKs}. Stickiness near the stability islands gives rise to subdiffusion, whilst the stickiness near the accelerator modes gives rise to the superdiffusion. Which of the structures prevails in the diffusion process is difficult to predict, as it depends on the details of such structures.

Our methods of analysis include the GALI method to characterize the structure of regular and chaotic regions, and also to quantify the degree of chaoticity, in a much better way than the ordinary Poincar\'e maps. The other measure of statistical and diffusive behavior is the diffusion exponent $\mu$, which we calculate both, by taking an average over a large number of initial conditions spread uniformly over the entire phase space  $(\theta,P)\in [0,2\pi]\times [0,2\pi]$, and also by taking an average inside the small cells, again with many initial conditions. The ``landscape'' of $\mu$ clearly correlates with the ``landscape'' of GALI$_2$, including the tiny structures in the phase space. Along with the $\mu$ we also calculate the \textit{effective} diffusion constant $D_{\rm eff}$, which, however, does not convey as much information as the $\mu$ itself, but its dependence on $K$ exhibits well known oscillations associated with the normal diffusion, and the peaks growing with time are positioned at accelerating modes. We do this analysis for several different values of $K$.

The central result of this paper is Fig.~\ref{figKvsmu}, where we plot $\mu$ as a function of $K$, which is complex, well converged (increasing the number of iterations does not change the graph), and clearly reveals  the association of the superdiffusion with the existence of the accelerator modes, even if the
average is taken over the entire phase space and the accelerator modes are relatively small. For $K> 2\pi$ the accelerator modes are largely of period one, whilst for $K<2\pi$ they are of higher period, and their influence on $\mu$ decays with $K$ much faster than for period one accelerator modes.

In order to see more details behind the superdiffusion at the peaks of the $\mu(K)$ plot of Fig.~\ref{figKvsmu}, we have also looked at the distribution function of the momentum $P$ after a large number of iterations, and taking a large number of initial conditions inside the starting cell. The result is
that we observe generally always an $\alpha$-stable L\'evy distribution, ``equipped'' with two ballistic peaks at plus or minus large $P$ value. These two peaks are directly produced by the initial conditions {\em inside} the accelerator modes. If we ignore them, that is cut them out, and renormalize the distribution, and fit it, it is always found to be the L\'evy distribution.

We do find that for the values of $K$ between the peaks of the $\mu(K)$ curve, and for the great majority of the initial conditions, the behavior is like for ordinary random walk, having the Gaussian distribution of the momenta (the special case of the L\'evy distribution with $\alpha=2$), and its variance growing linearly with time, $\mu=1$.

In conclusion, we may emphasize that all features of anomalous diffusion in the standard map can be understood in terms of the accelerator modes, and the sticky objects surrounding them in the form of cantori, as well as around the regular islands. Moreover, all these features can be quantified by the methods that we employ.

The main motivation for the present work was our recent research on the quantum kicked rotator \cite{ManRob2013PRE,BatManRob2013}, which is the quantized
classical kicked rotator, described by the standard map. For the systematic understanding of the quantum kicked rotator and its semiclassical theory, a complete survey of the standard map is necessary. We have achieved that by going as far as $K \le 70$, where it is expected that for $K>70$ there are essentially no new features, as $\mu(K)$ is almost everywhere equal to 1, except within narrow intervals containing the accelerator modes of period 1. As for the association of the quantum localization and the classical diffusion, the semiclassical theories  must be improved, although the first step has been undertaken in \cite{ManRob2013PRE}. On the classical side, the theory and derivation of the diffusion exponent $\mu$, and other aspects of anomalous diffusion, are still lacking and open for the future.\\

\section*{Acknowledgements}
This work was supported by the Slovenian Research Agency (ARRS). T.M. was also partially supported by a grant from the Greek national funds through the Operational Program ``Education and Lifelong Learning'' of the National Strategic Reference Framework (NSRF) - Research Funding Program: THALES. Investing in knowledge society through the European Social Fund.


\bibliography{ManRobPRE}

\end{document}